\documentclass[preprint,12pt]{elsarticle}
\usepackage{graphics}
\usepackage{hyperref}
\usepackage{xfrac}
\usepackage{graphicx} 
\usepackage[font=small,labelfont=bf]{caption} 
\usepackage[subrefformat=parens]{subcaption}
\usepackage{amssymb}
\usepackage{multirow}
\usepackage[LGRgreek]{mathastext}

\begin{document}

\begin{frontmatter}

\title{Simulation of heat transfer and dissipation in targets used in nuclear astrophysics experiments}

 \author[label1]{Tanmoy Bar}
 \author[label1]{Chinmay Basu}
 \author[label2]{Mithun Das}
 \author[label2]{Apurba Kumar Santra}
 \author[label3]{Swarnendu Sen}
 \address[label1]{Saha Institute of Nuclear Physics, HBNI, 1/AF, Bidhannagar, Kolkata-700064}
 \address[label2]{Department of Power Engineerining, Jadavpur University, Salt Lake Campus, Sector- III,  Kolkata- 700098}
 \address[label3]{Department of Mechanical Engineering, Jadavpur University, Kolkata- 700032}

\begin{abstract}

This work presents time-dependent numerical calculations of heat generation and dissipation in targets used in high ion-beam current nuclear astrophysics experiments. The simulation is beneficial for choosing the thickness of targets, maximum ion-beam current and design setup for cooling of such targets. It is found that for the very thin target ($^{27}Al(p,p),^{12}C(p,p)$) heat generation inside target is relatively low and a fair amount of high current (few $\mu$A)can be used without any melting issue. But in case of thick targets ($^{27}Al(p,\gamma),^{12}C(p,\gamma)$) cooling became essential for the survival of reaction target.

\end{abstract}
\begin{keyword}
High current ion-beam, targets, heat dissipation, temperature profile.
\end{keyword}
\end{frontmatter}

\section{Introduction}
Stellar nucleosynthesis\cite{rolf1988cauldron} proceeds through low energy fusion or capture reactions with very low cross sections ($\sim$nb to pb). The measurements of these low cross-sections with appreciable accuracy is extremely difficult. In order to improve the statistical accuracy in the cross section, a high beam current with a thick target is useful. The typical beam current in nuclear astrophysics(NA) experiments can vary from a few mA to several 100$\mu$A. The targets may vary from low to high thermal conductivities. High beam currents however generate a large amount of heat in a solid thick target and if the temperature exceeds the melting point of the target material, a damage of the target is inevitable. The use of windowless gas or gas jet targets can avoid this problem, but they are expensive and require a quite elaborate setup. So a solid target will require a cooling mechanism so that the generated heat, may be dissipated to keep the temperature well below the melting point, and the target can be used over a reasonable beam time period. Other methods like rotating targets or beam wobbling are not in general useful for all type of experiments. As experiments may require a difficult setup for cooling, a detailed theoretical study is essential. Ther are some works \cite{greene2001temperature,rovais2011computer,
wang2015heat,tian2001target,guo2016experimental} that simulate heat generation in different types of targets, but a general study is absent.\\
\indent In this work we present a 3d heat transfer simulation systematically for targets used in NA experiments\cite{morales2015tests,chiari2001proton,harissopulos200027,mazzoni1998proton,burtebaev2008new,spillane2007c}. We considered Aluminium and Carbon as targets as these possess good and bad thermal conductivitites respectively. The results shows the limiting thickness and beam currents that can be used with and without cooling conditions.

\section{Mechanism of heat generation and dissipation in targets}
\subsection{PROBLEM FORMULATION}
\begin{figure}
  \centering
    \includegraphics[width=.5\linewidth]{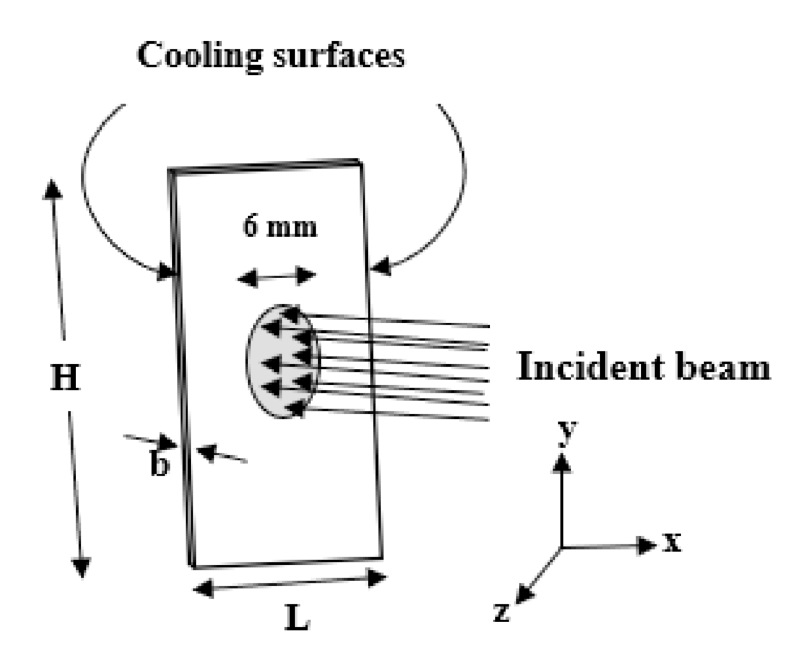}   
  \caption{Schematic picture of reaction target with cooling.}
\end{figure}
The schematic details of the expriment has been shown in Figure 1. Target has a dimension of H $\times$ L $\times$ b. Central circular region of 6mm diameter is exposed to the beam. This beam then generates heat inside the target. In some cases to cool down the target, cooling is used on two sides (as shown in Figure 1). When this cooling is not sufficient cooling from back ( only for $\gamma$-detection experiments) is also considered. Here simulation has been carried out to know the limiting target thickness and beam current. 
\subsection{GOVERNING EQUATIONS}

Calculation of heating in these targets will give a guideline to the safe maximum beam currents they may be exposed before melting.Calculation is taken into account the amount of heat the beam produces inside the target and how that heat is dissipated through conduction and radiation process.The balanced equation is:
\begin{center}
    Heat in = Heat out
\end{center}
This equation needs to be satisfied to reach a steady state. Heat is brought into account inside the target by the energy loss of the beam inside the foil(target). This may be calculated by using a stopping power model such as SRIM\citep{ziegler2010srim}. Under fixed conditions heat production inside the foil is proportional to the beam current. The heat in the target gradually decreases over time by conduction through the target away from the beam spot and radiation, which is given by Stefan-Boltzmann law as:
\begin{equation} E=\epsilon\sigma S(T^4 - T_0 ^4)
\end{equation}
where E is the radiant heat energy emitted per unit time; $\epsilon$, emissivity of the target material; $\sigma$, Stefan-Boltzmann constant ($\sim 5.67\times 10^{-8} watt/m^2 . K^4$); S is the surface area irradiated by the Gaussian shaped beam and $T_0$ is the ambient temperature of the target surrounding. Now to have an idea about steady state temperature one need to solve three dimentional heat conduction equation\citep{bergman2011fundamentals} with boundary conditions.
\begin{equation}
\nabla ^2 T = \frac{1}{\alpha} \frac{\partial T}{\partial t}
\end{equation}
where $\alpha = \frac{k}{\rho C_v}$. In case of very thin targets calcualtion has been done two dimentionally because then is not much change in temperature profile due to very thin targets.\\ Boundary conditions used to solve this equation are follows [see Fig.1]:\\
At x = 0 and L, T = $T_c$  where $T_c$, cooling temperature.\\
at y = 0 and H final temperature is given by Eq. 1\\
at z = 0 also final temperature is given by Eq. 1\\
at z = W for central circular area temperature generated by heat flux (Eq.4) otherwise initial temprature is $T_0$ and changes according to Eq. 1.\\
Steady state numerical calucation has been done in\cite{greene2001temperature} where they used equation like below to find steady state temperature.
\begin{equation}
    WI=mC_v \frac{dT}{dt}+(T-T_0)\frac{\lambda S}{D/\rho}+2\epsilon\sigma S(T^4 - T_0 ^4)    
\end{equation}
where left hand side is the amount of heat generated by the incident beam in the target and  right hand side is the different processes to dissipate that energy. Here temperature, T is a function of x, y, z. W equals to the energy loss in the target by each projectile,I is the number of projectile coming per unit time. Which is generally expressed as particle current, m and $C_v$ are mass and the specific heat of target, $T_0$ is the ambient temperature, $\lambda$ is the thermal conductivity of the target,
S is the surface area of the target, $\rho$ is the density of target material,
 D is the areal thickness.\\
Here heat simulation package ANSYS has been used to find the steady state temperature and temperature profile of the target. 

\subsection{CALCULATION OF HEAT FLUX}

All heat dissipation calculations in ANSYS program is done by putting equivalent heat generated inside target material for real nuclear astrophysics reactions as a source term. To calculate the amount of heat generated inside the target foil for a particular beam current we used following formula:
\begin{equation}
     H = \textit{i} ^p \times R \times \textit{Stopping power}\\
     \end{equation}
     Here, H is the amount of heat generated inside the target per sec (watt),
     ${\textit{i}}^p$ is the particle current of projectile and R is the range of the projectile inside that target. Now the heat flux, $\mathcal{H}$ is given by,
     \begin{equation}
     \mathcal{H} = \frac{H}{A}
\end{equation}
where A is the beam spot area.

\section{Numerical method}
For numerical study ANSYS software has been used. Finite element method has been used to solve the problem. The geometry is constructed by ANSYS spaceclaim and then the governing equations has been discretised to solve it numerically considering the given boundary conditions. Transient thermal module has been used to determine the steady state temperature distribution and heat propagation period through the target. Details of this can be found in official website\cite{WinNT} of ANSYS. A grid independence test has been carried out to find the number of element which gives optimum accuracy. Increasing the number of element is not improving the output appreciatively but increases computational time and space required. Details of grid independent test has been shown in Figure 2.

 \begin{figure}
  \centering
    \includegraphics[width=.5\linewidth]{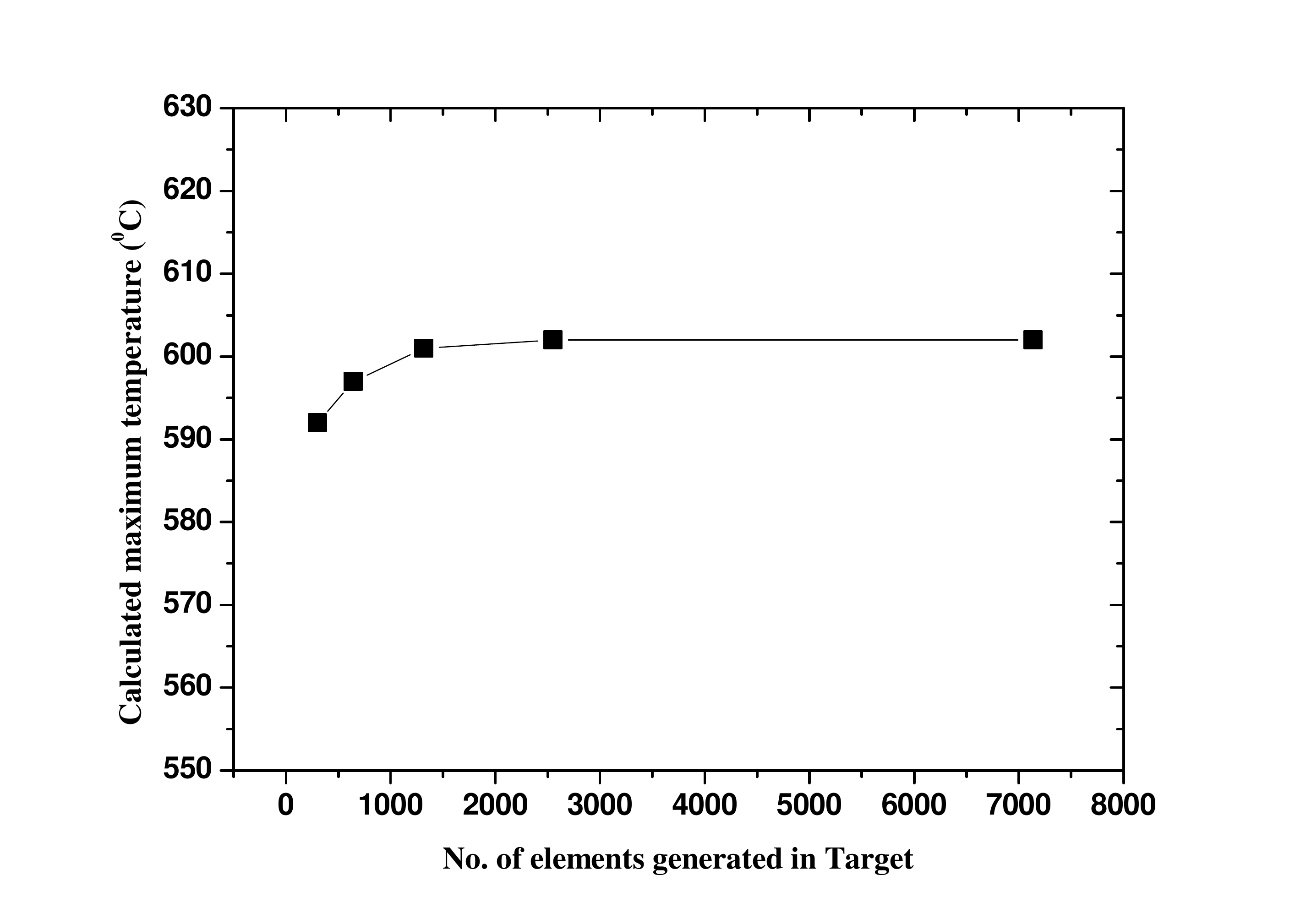}
    \caption{Grid independency test curve}
  \end{figure}
Above grid indepent test has been done for $^{27}Al(p,p)$ reaction study. Same has been done for each case to have grid independent solution. For this particular case number of elements over 2500 gives same results so we have taken around that number to solve our problem. 

\section{Results and discussions}

The simulation was carried out for three types of reactions which are most common in the field of NA. These are the proton scattering experiments viz. (p,p), proton capture(p,$\gamma$) and $^{12}C(^{12}C,x)$. The inputs for the calculations for these reactions were obtained from the literature. $^{27}Al$ and $^{12}C$ targets were used in these experiments. Our aim is to find the maximum permissible beam current for a particular target thickness and vice-versa. All calculations were done with a sample foil size of 25mm$\times$25mm cross sectional area and beam spot diameter of 6mm. The ambient temperature in which the target beam interactions took place in absense of any external cooling was considered to be $16^oC$. Few cases has been studied and parameters for those studies has been listed in TABLE 1.
\begin{table}
\begin{center}
\begin{tabular}{|p{2cm}||p{1cm}|p{2cm}|p{2.5cm}|p{1.8cm}|p{1.7cm}|}
\hline
Reaction & Beam energy in MeV & Beam current & Target thickness &Stopping power in $\frac{MeV}{(mg/cm^2)}$  & Heat flux in W/$m^2$ \\ \hline
\multirow{2}{*}{$^{27}Al(p,p_0)$} & \multirow{2}{*}{1.1} & 15$\mu$A & \multirow{2}{*}{39 $\mu g/cm^2$} &\multirow{2}{*}{0.1654} &$3.42\times 10^3$\\\cline{3-3}\cline{6-6}
 &&16$\mu$A &&& $3.65\times 10^3$\\
\hline \hline
\multirow{3}{*}{$^{27}Al(p,\gamma)$} & \multirow{3}{*}{0.8} & 3$\mu$A & \multirow{2}{*}{270.2 $mg/cm^2$} &\multirow{3}{*}{0.1654} & $5.9\times 10^4$\\\cline{3-3}\cline{6-6}
 &&650$\mu$A &&& $1.28\times 10^7$\\\cline{3-4}\cline{6-6}
  &&7$\mu$A &2.702$mg/cm^2$&& $1.37\times 10^5$\\
\hline \hline
\multirow{2}{*}{$^{12}C(p,p)$} & \multirow{2}{*}{0.35} & 40nA & \multirow{2}{*}{13 $\mu g/cm^2$} &\multirow{2}{*}{0.4385} &8.06\\\cline{3-3}\cline{6-6}
 &&5mA &&& $1.0\times10^6$\\
\hline \hline
\multirow{2}{*}{$^{12}C(p,\gamma)$} & \multirow{2}{*}{0.5} & 15$\mu$A & \multirow{2}{*}{1338 $mg/cm^2$} &\multirow{2}{*}{0.1646} & 2.13$\times10^5$\\\cline{3-3}\cline{6-6}
 &&3.5mA &&& $4.97\times10^7$\\
 \hline \hline
 \multirow{3}{*}{$^{12}C(^{12}C,x)$} & \multirow{3}{*}{4.5} & 40$\mu$A & \multirow{2}{*}{225.3 $mg/cm^2$} &\multirow{3}{*}{7.346} & $4.0\times10^6$ \\\cline{3-3}\cline{6-6}
 &&250$\mu$A &&& $2.51\times10^7$\\ \cline{3-3}\cline{6-6}
 &&1mA &&& $1.0\times10^8$\\
\hline 
\end{tabular}
\caption{Different parameter values used in simulation. (For $^{12}C(p,\gamma)$, values are taken for Cu backing) }
\end{center}
\end{table}

\subsection{ $^{27}Al(p,p_0)$ REACTION:}
This reaction was studied by M.Chiari et al.  \cite{chiari2001proton} where the target thickness was 39 $\mu$g/$cm^2$ and the beam current was varried between 50-150nA. The energy of the proton beam was 1.1 MeV. We considered two cases; (i) where the cooling is only through radiation, (ii) where the cooling is induced by chillled water on the two sides of the target. We do not consider any other mounting frames associated with target. Figure 3(a) shows the temperature profile and in 3(b) the temporal variation of the temperature. In Figure 4(a) and 4(b) we show the same with external cooling by chilled water ($4^oC$) at the two sides. The maximum current attained before melting is found to be 15$\mu$A. Though there is a change in the temperature distribution due to the cooling, the maximum current attained is not much altered (16$\mu$A). Since area at two sides of thin target is very less, the heat dissipation through chilled water is less effective.
\begin{figure}
  \centering
  \begin{subfigure}[b]{0.48\linewidth}
    \includegraphics[width=\linewidth]{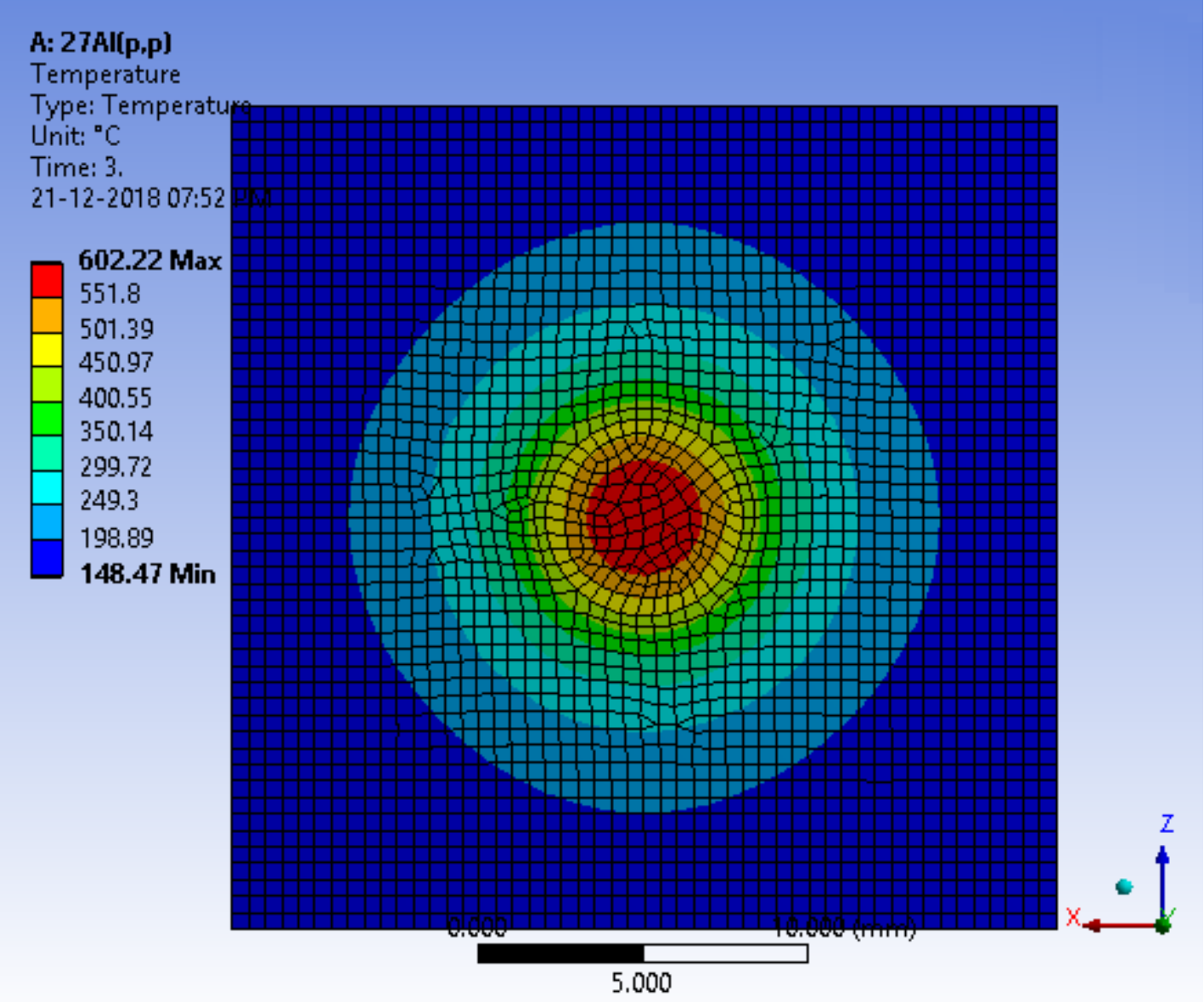}
    \caption{Temperature profile}
  \end{subfigure}
  \begin{subfigure}[b]{0.48\linewidth}
    \includegraphics[width=\linewidth]{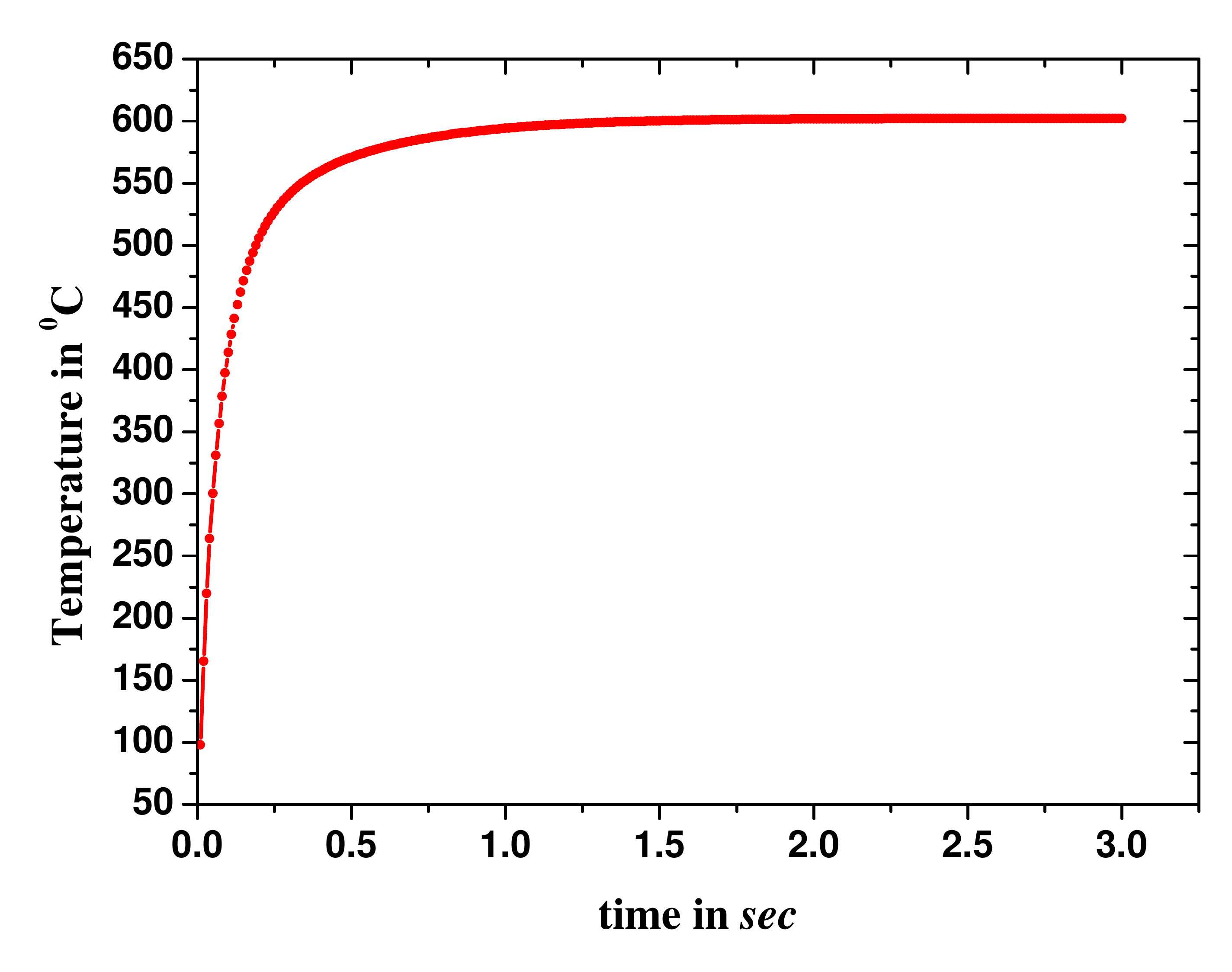}
    \caption{Temporal variation of maximum temperature}
  \end{subfigure}
  \caption{Temperature study of $^{27}Al(p,p_0)$ reaction without any cooling system applied.}
\end{figure}

\begin{figure}
  \centering
  \begin{subfigure}[b]{0.48\linewidth}
    \includegraphics[width=\linewidth]{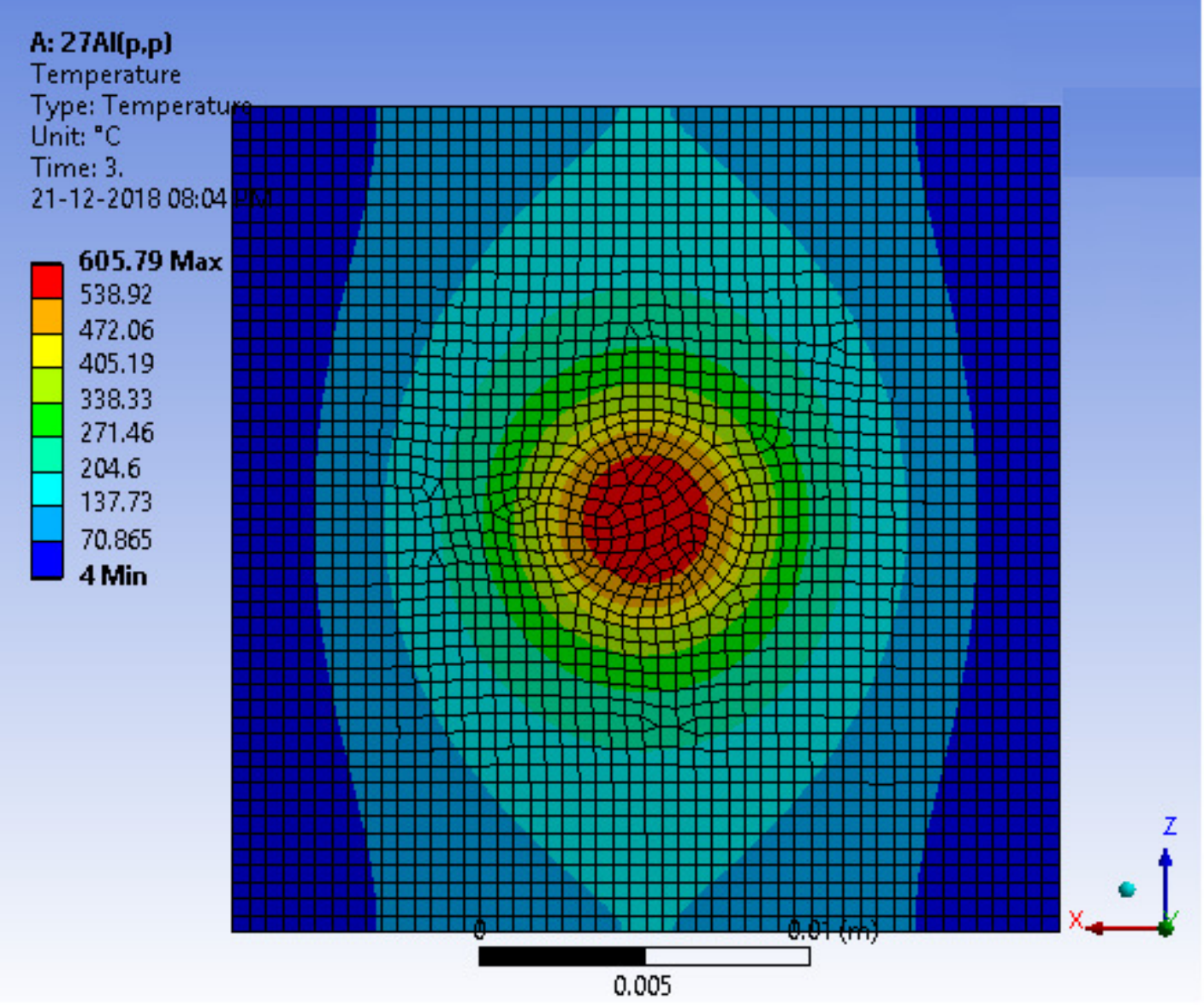}
    \caption{Temperature profile}
  \end{subfigure}
  \begin{subfigure}[b]{0.48\linewidth}
    \includegraphics[width=\linewidth]{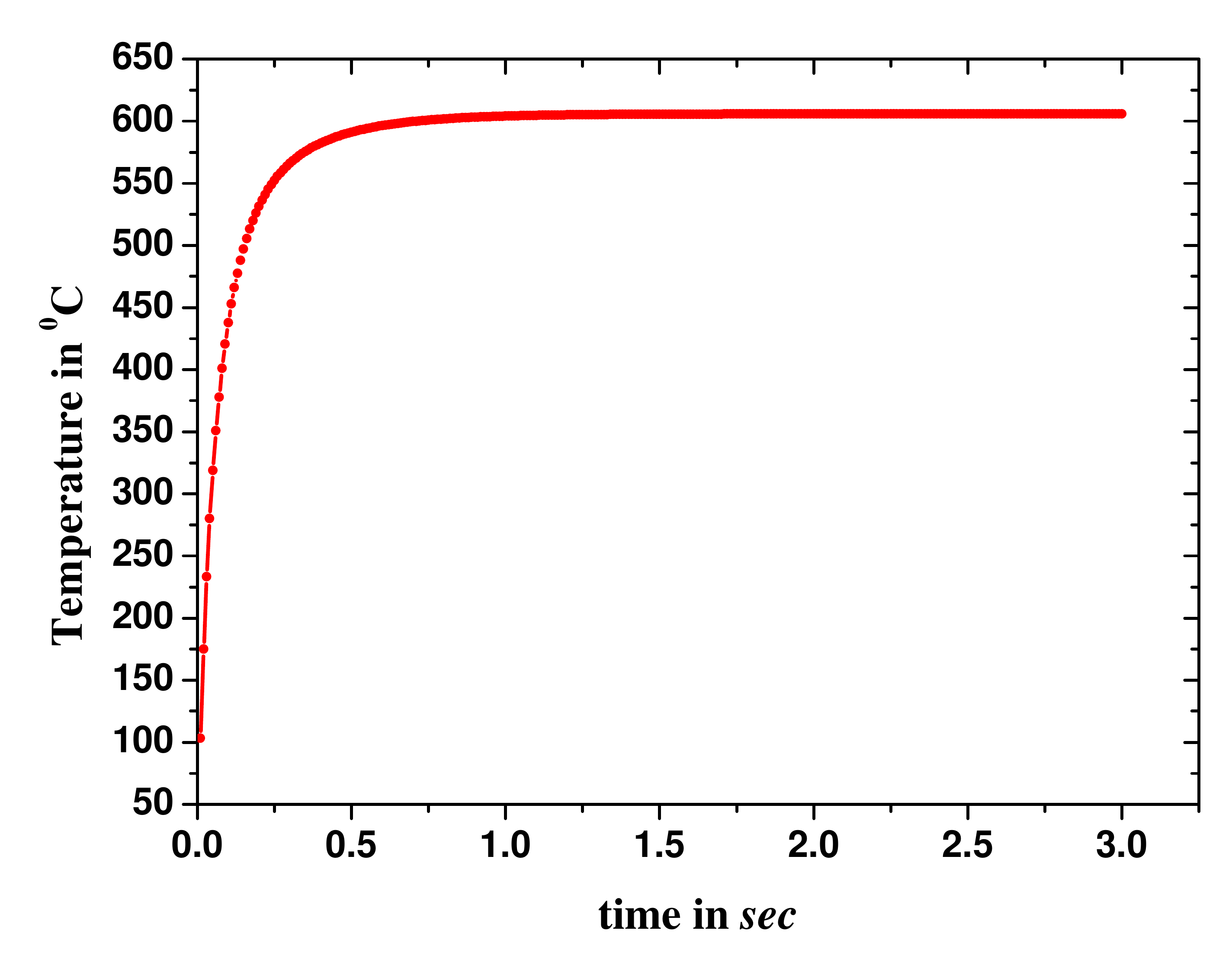}
    \caption{Temporal variation of maximum temperature}
  \end{subfigure}
  \caption{Temperature study of $^{27}Al(p,p_0)$ reaction with two side $4^o$C cooling system applied.}
\end{figure}
\subsection{$^{27}Al(p,\gamma)$ REACTION:}
The $^{27}Al(p,\gamma)$ reaction was carried out using a 45 $\mu$g/$cm^2$ target at proton energy of 0.8 MeV \citep{harissopulos200027}. The beam current was about 600nA. The results with dissipation by only radiation and with external cooling as well as radiation as in Sec 4.1 are shown in Figure 5 and 6. For 600nA proton beam with energy 0.8 MeV, target of any thickness can be used. Table 2 shows some maximum saturation temperature for various target thicknessess. It shows that as thickenss is increased maximum temperature decreases, this is happening because of the increase in radiation area. For a 1mm thick target maximum of 3$\mu$A can be used without any cooling[Figure 5]. Operating current beam can be increased to 650$\mu$A after applying two sides cooling by $4^0$C [Figure 6]. Now decrease in target thickness also decreses the cooling area, so for thin target of thickness 0.01mm (2.7$mg/cm^2$) beam current is restricted to 7$\mu$A [Figure 7].

\begin{table}[h!]
\begin{center}
\begin{tabular}{ |p{0.3cm}||p{2cm}||p{2cm}||p{2.5cm}|  }
 \hline
 Sl. & Target thickness in mm & Target thickness in $mg/cm^2$ & Maximum steady state temperature in $^0$C  \\ \hline \hline

 1&     0.02    &   5.404 &   339 \\  \hline
 2&     0.1     &   27.02 &   332 \\  \hline
 3&     1       &   270.2 &   310\\  \hline
 4&     4       &   1080.8 &  283\\  \hline
 
\end{tabular}\\
\vspace{0.2cm}

\caption{Results for various thickness of Al target for 0.8 Mev proton beam fixed at 600nA.}
\end{center}
\end{table}

\begin{figure}
  \centering
  \begin{subfigure}[b]{0.48\linewidth}
    \includegraphics[width=\linewidth]{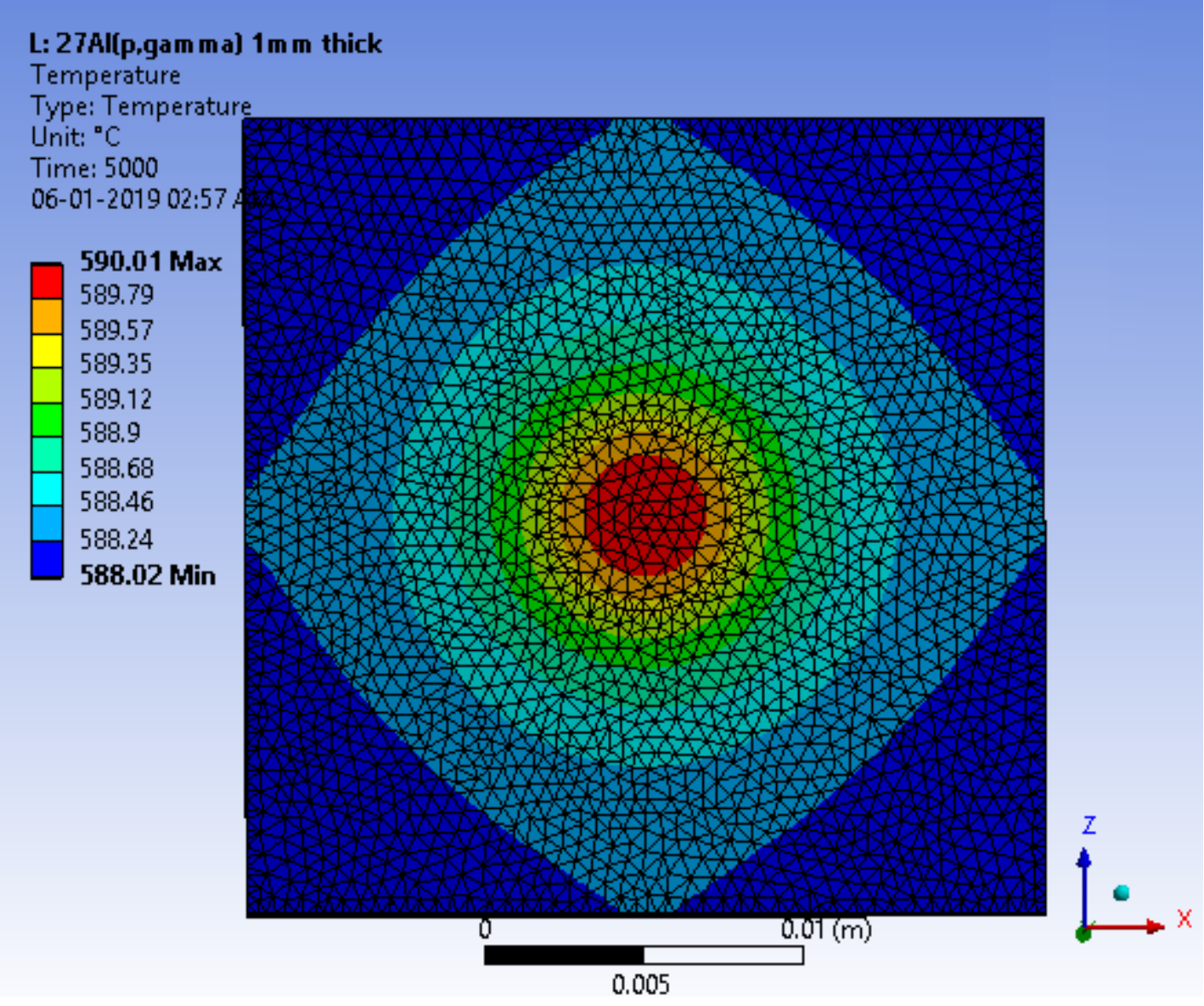}
    \caption{Temperature profile}
  \end{subfigure}
  \begin{subfigure}[b]{0.48\linewidth}
    \includegraphics[width=\linewidth]{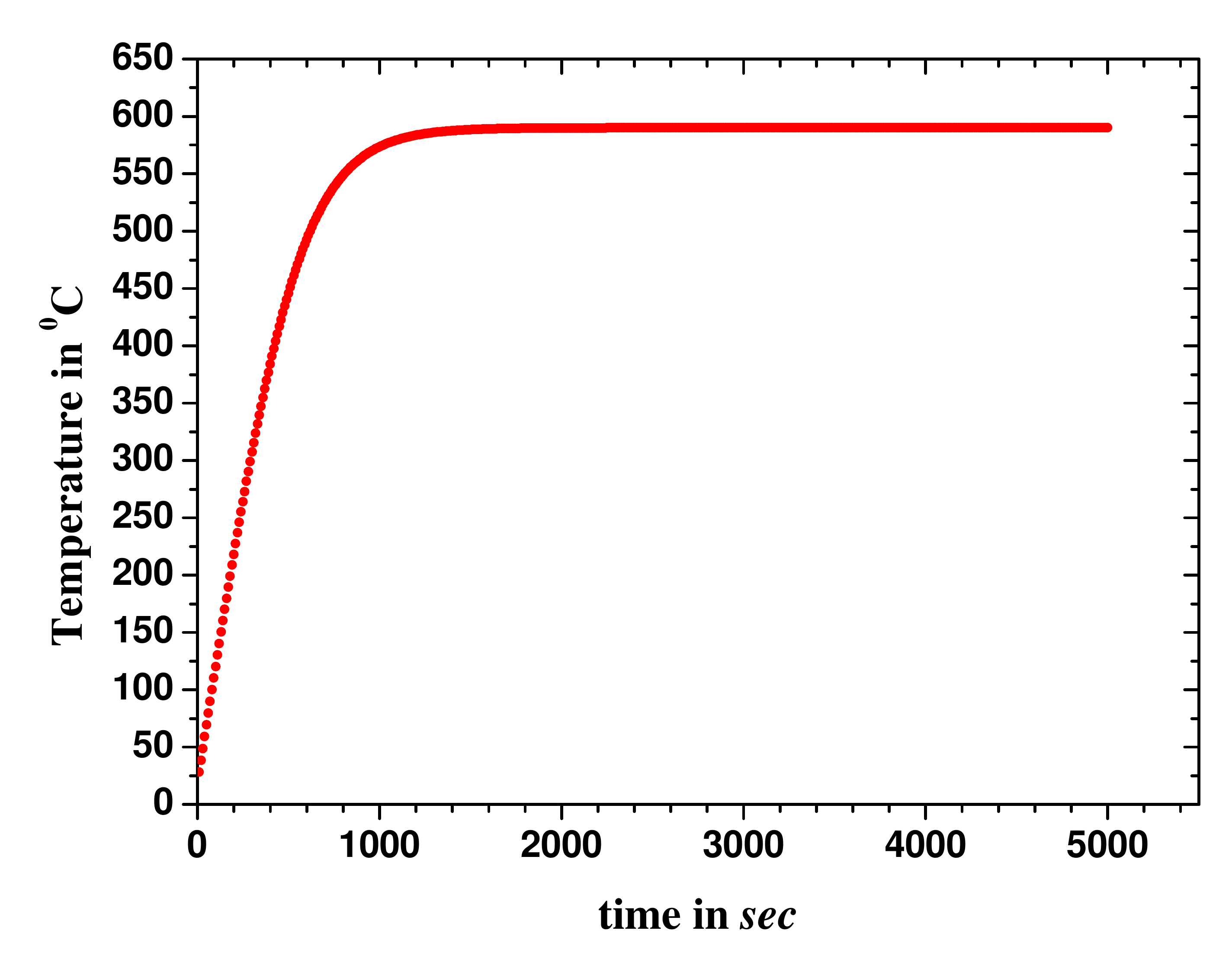}
    \caption{Temporal variation of maximum temperature}
  \end{subfigure}
  \caption{Temperature study of $^{27}Al(p,\gamma)$ reaction without any cooling for 1mm target.}
\end{figure}

\begin{figure}
  \centering
  \begin{subfigure}[b]{0.48\linewidth}
    \includegraphics[width=\linewidth]{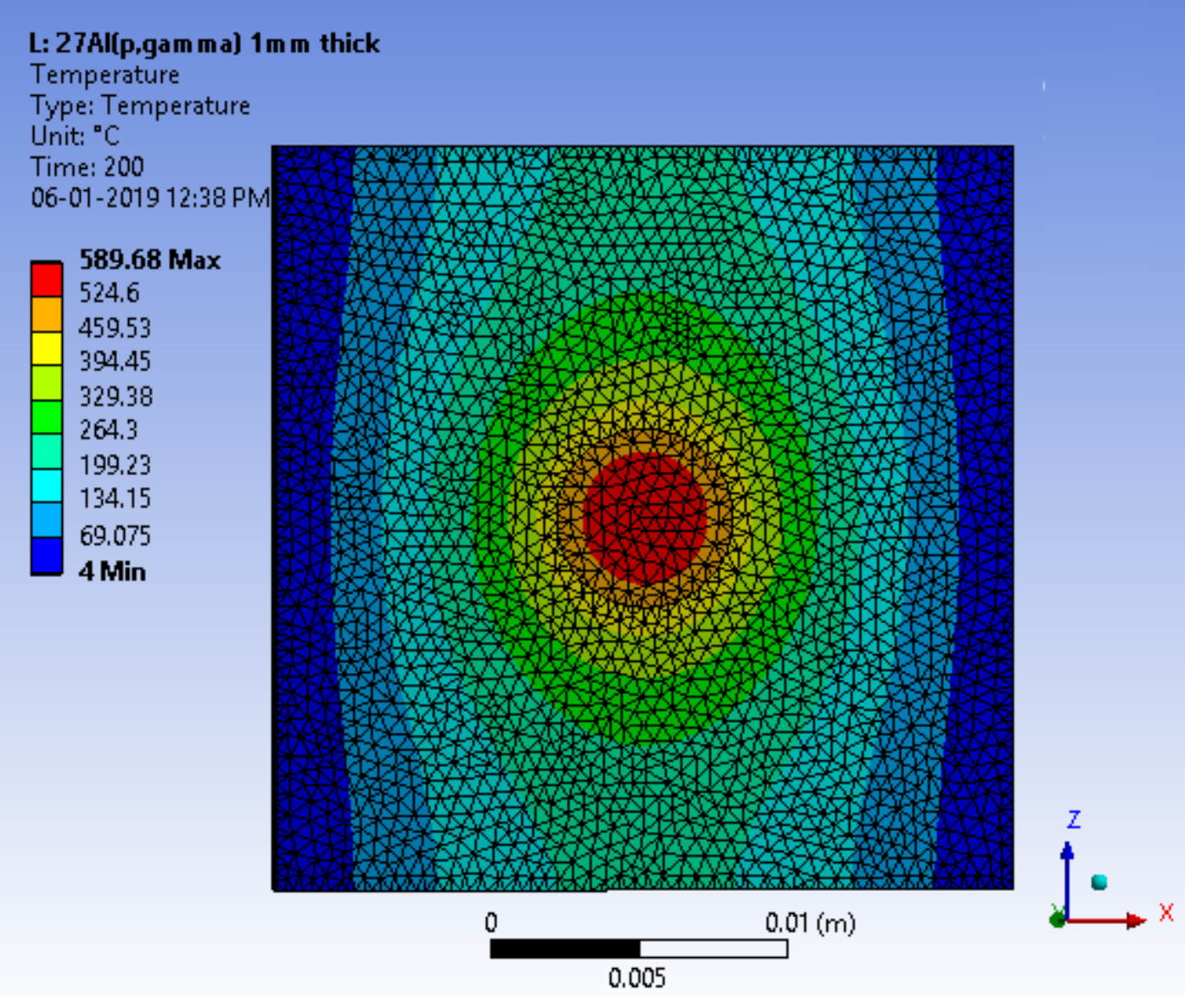}
    \caption{Temperature profile}
  \end{subfigure}
  \begin{subfigure}[b]{0.5\linewidth}
    \includegraphics[width=\linewidth]{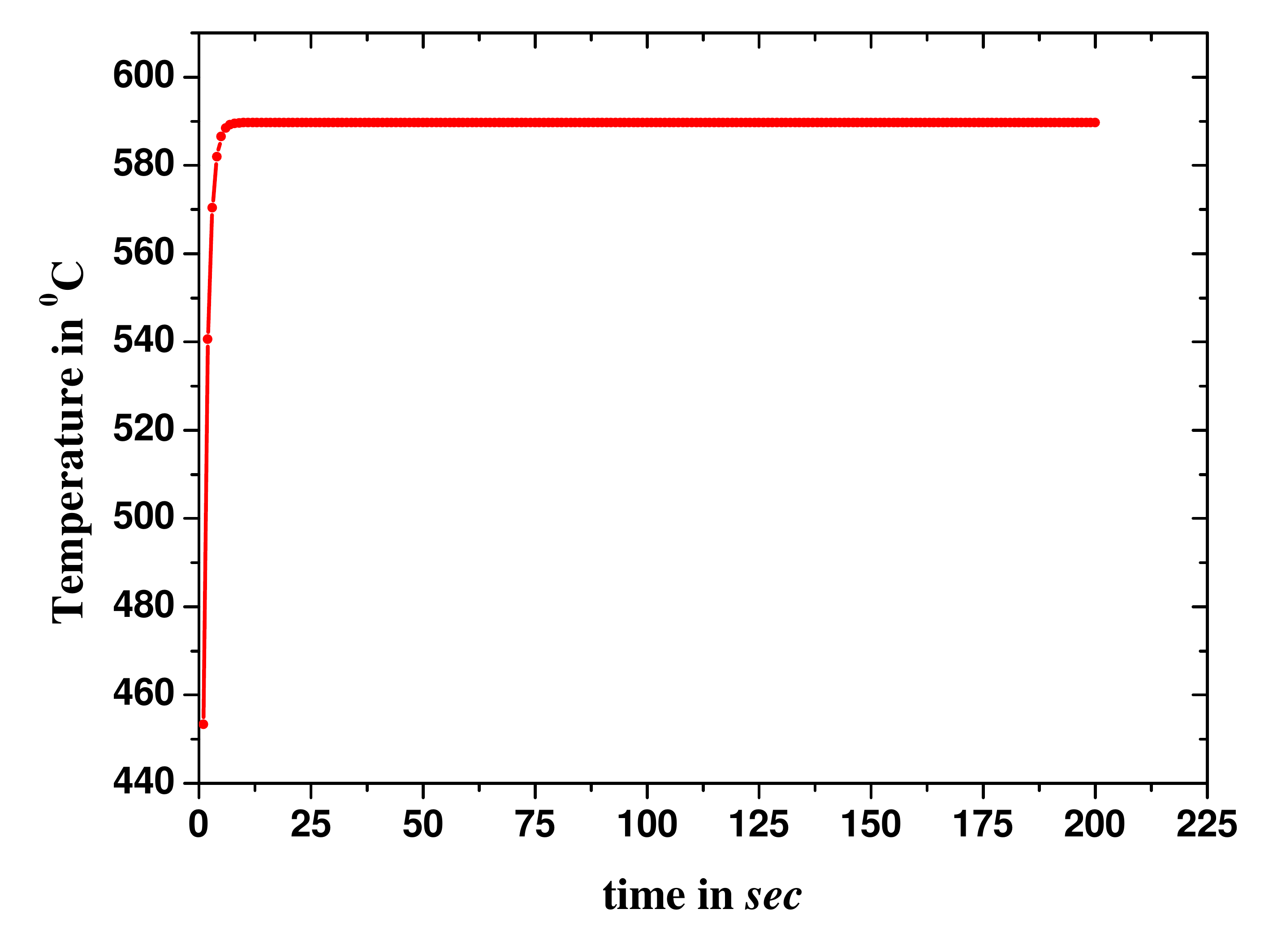}
    \caption{Temporal variation of maximum temperature}
  \end{subfigure}
  \caption{Temperature study of $^{27}Al(p,\gamma)$ reaction with two side $4^o$C cooling applied for beam current 650$\mu$A and thickness 1mm.}
\end{figure}

\begin{figure}
  \centering
  \begin{subfigure}[b]{0.48\linewidth}
    \includegraphics[width=\linewidth]{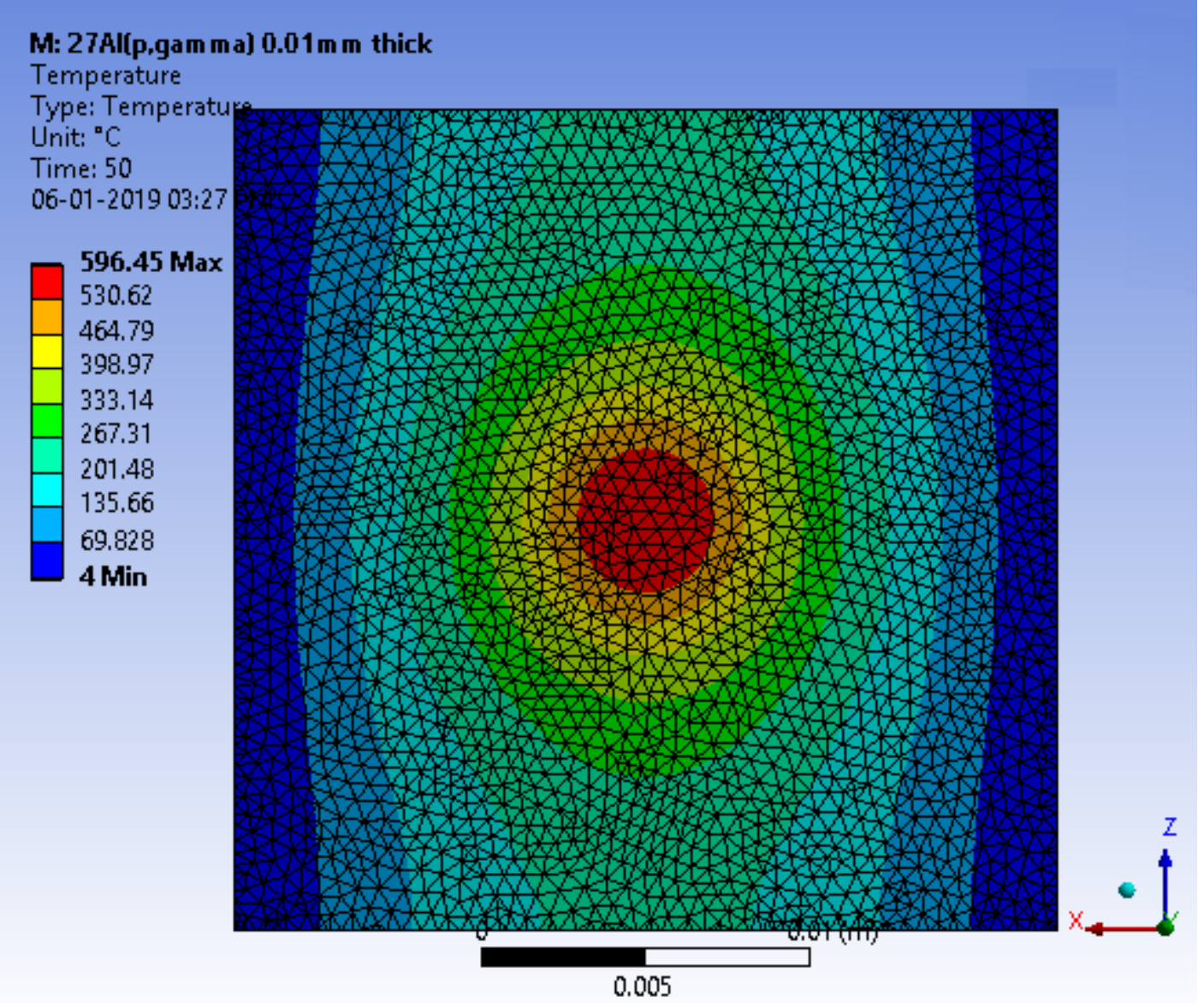}
    \caption{Temperature profile}
  \end{subfigure}
  \begin{subfigure}[b]{0.48\linewidth}
    \includegraphics[width=\linewidth]{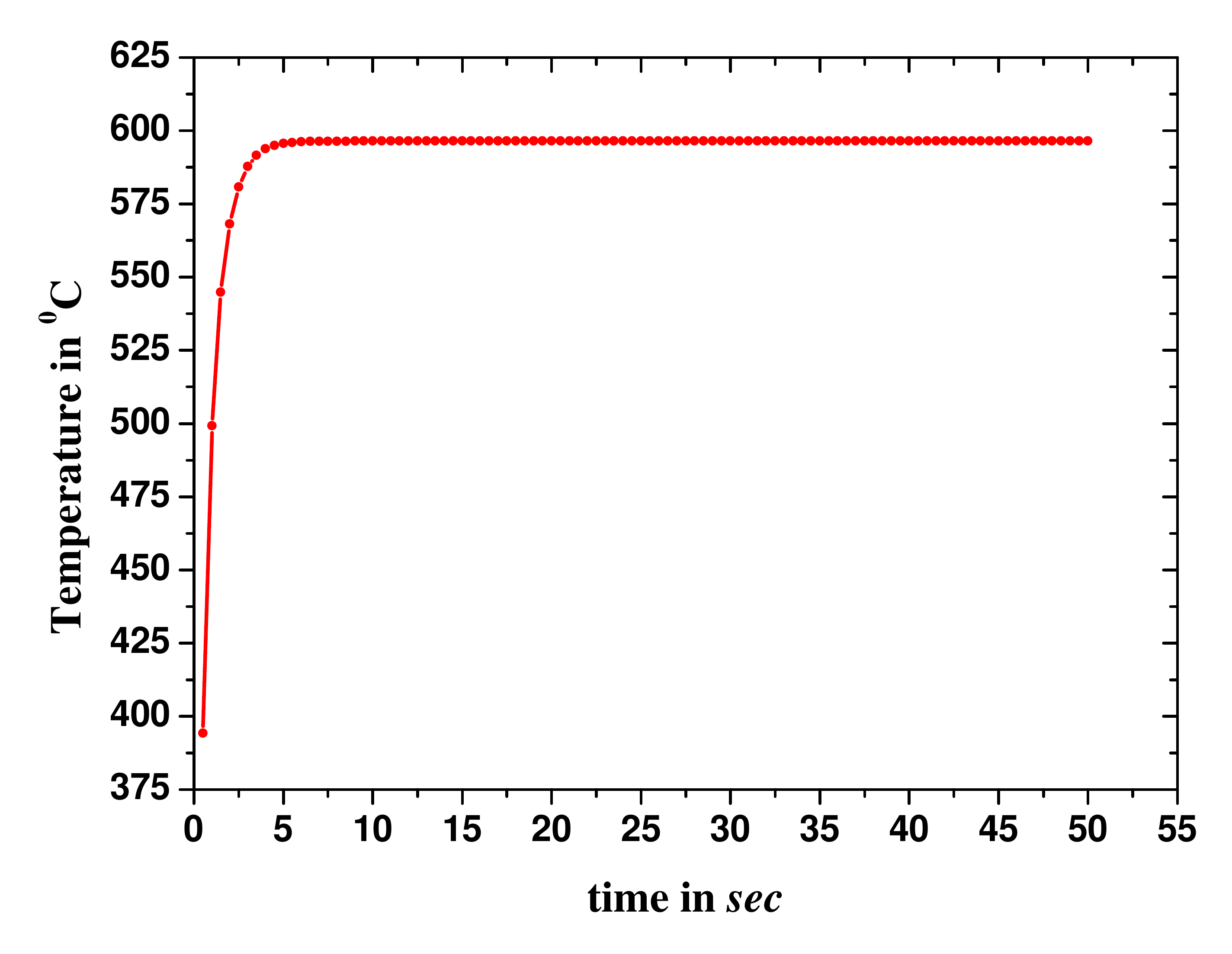}
    \caption{Temporal variation of maximum temperature}
  \end{subfigure}
  \caption{Temperature study of $^{27}Al(p,\gamma)$ reaction with two side $4^o$C cooling applied for 0.01mm thick target at 7$\mu$A beam current.}
\end{figure}
 \subsection{$^{12}C(p,p)$ REACTION:}
The experiment of S. Mozzoni et al.\citep{mazzoni1998proton} is considered for this reaction where 13 $\mu$g/$cm^2$
 To study $^{12}C(p,p)$ we used 13 $\mu$g/$cm^2$ graphite target was used with a beam energy of 0.35 MeV. The current used in the experiment was 40nA. The results with these parameters are shown in Figure 8(a) and (b). As the target is very thin the maximum temperature attained in the experiment is found from calculation to be around $18^oC$. A maximum of 5mA or even more can be used for this thin target without melting it. The steady state temperature fro 5mA was found to be $\sim 1900^oC$(Figure 9(a) and (b)), where cooling done by radiation only. Since there is no effect of cooling with thin targets, that calculation was not done.

\begin{figure}
  \centering
  \begin{subfigure}[b]{0.48\linewidth}
    \includegraphics[width=\linewidth]{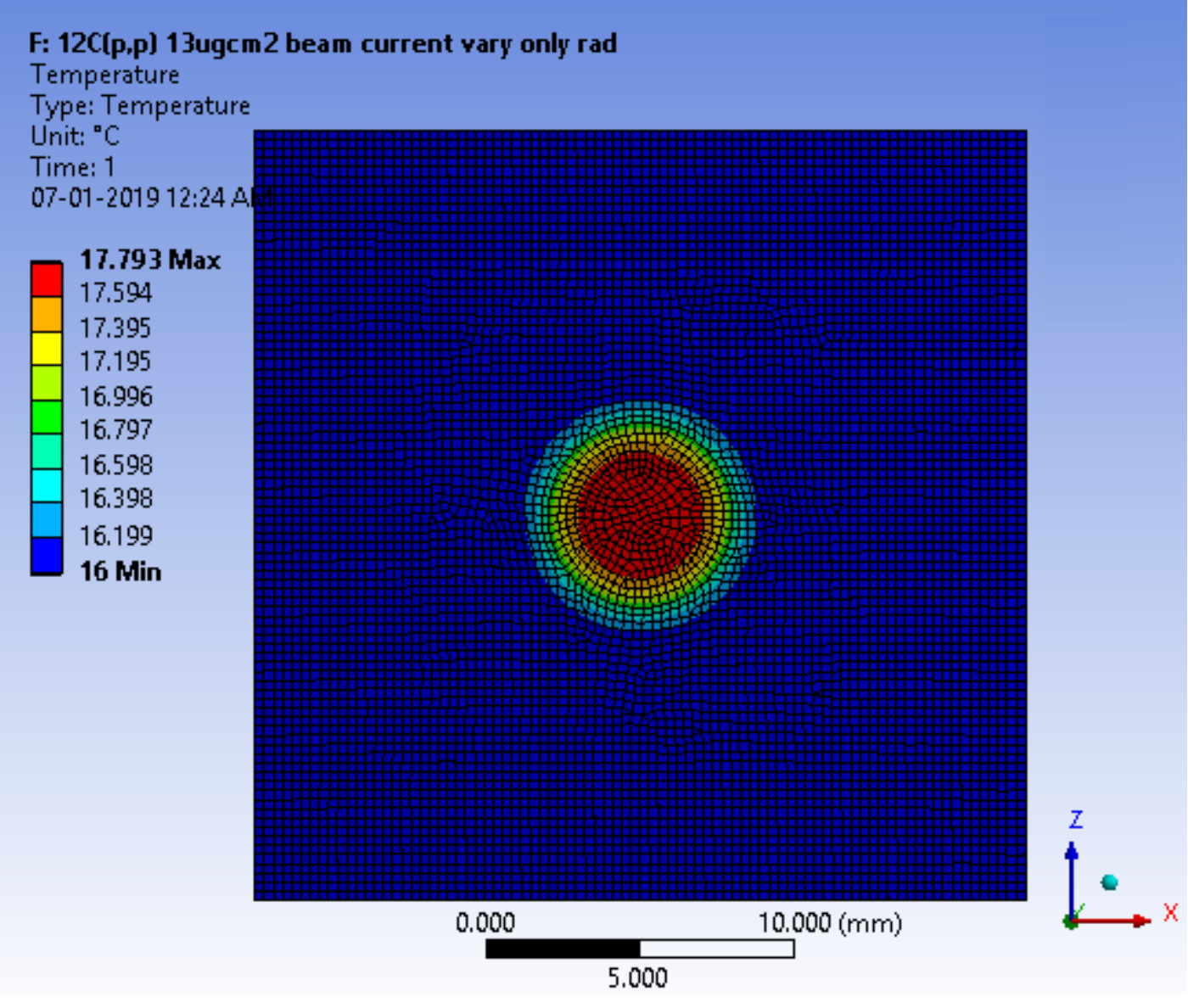}
    \caption{Temperature profile}
  \end{subfigure}
  \begin{subfigure}[b]{0.48\linewidth}
    \includegraphics[width=\linewidth]{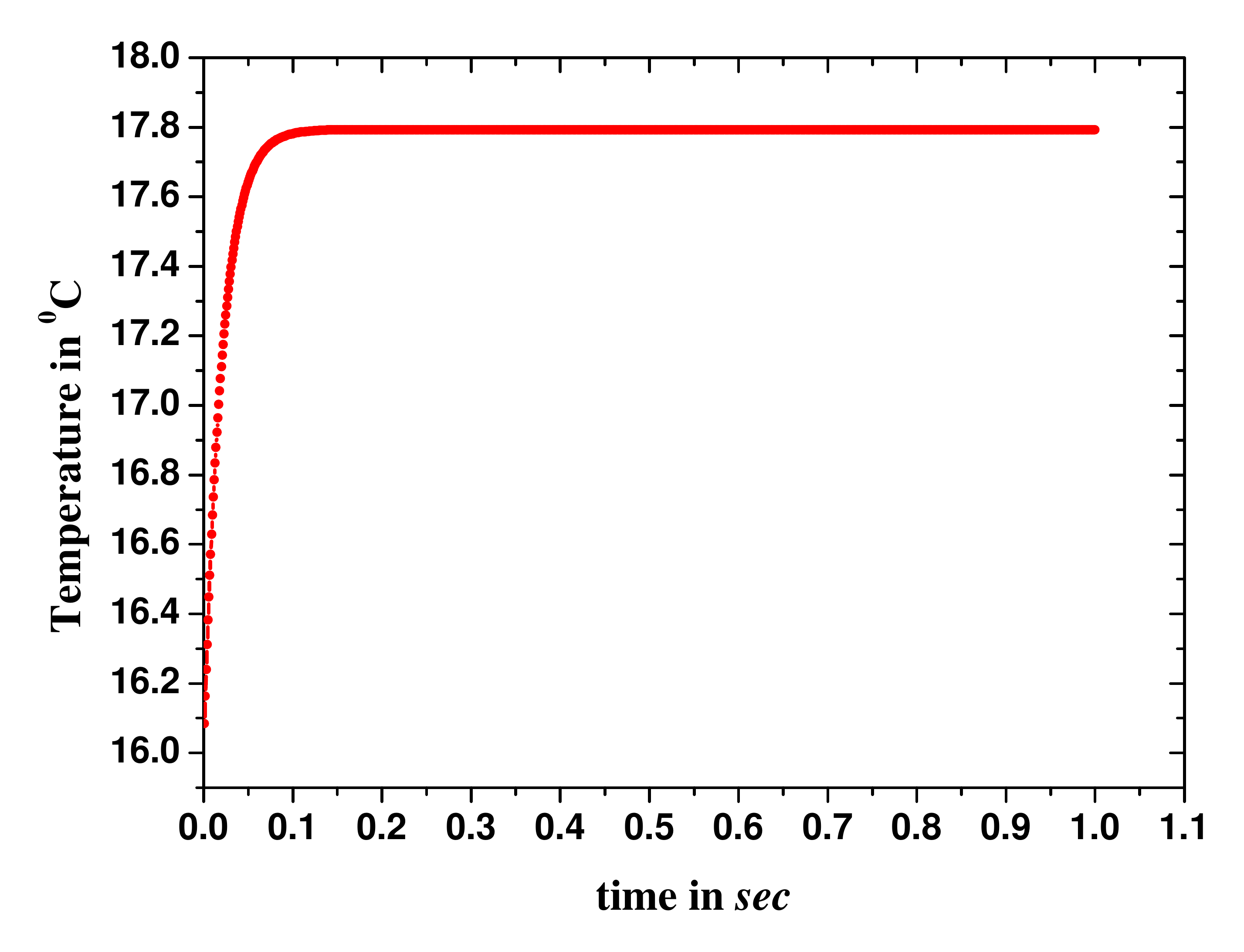}
    \caption{Temporal variation of maximum temperature}
  \end{subfigure}
  \caption{Temperature study of $^{12}C(p,p)$ reaction without any cooling for 40nA beam.}
\end{figure}

\begin{figure}
  \centering
  \begin{subfigure}[b]{0.48\linewidth}
    \includegraphics[width=\linewidth]{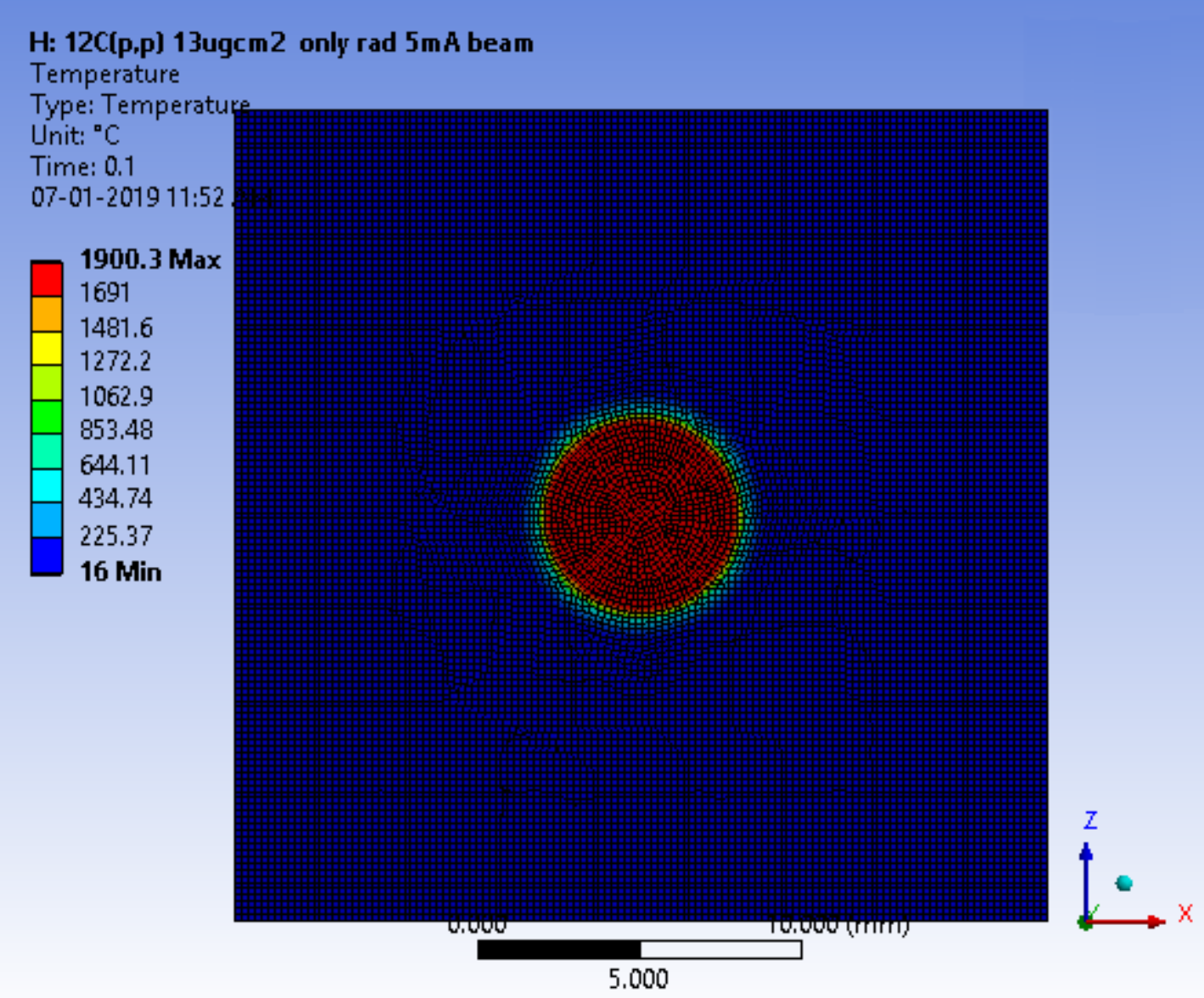}
    \caption{Temperature profile}
  \end{subfigure}
  \begin{subfigure}[b]{0.48\linewidth}
    \includegraphics[width=\linewidth]{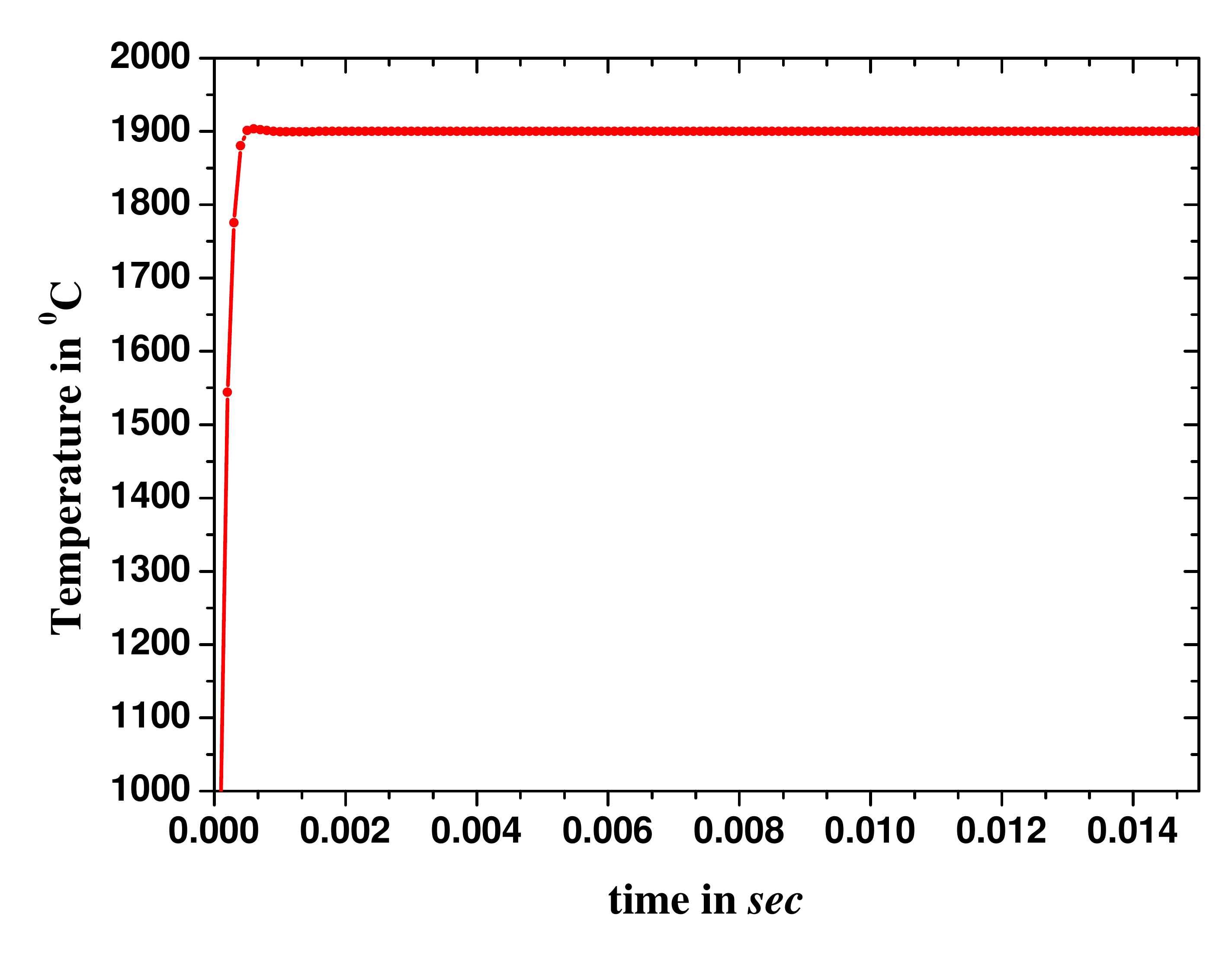}
    \caption{Temporal variation of maximum temperature}
  \end{subfigure}
  \caption{Temperature study of $^{12}C(p,p)$ reaction without any cooling arrangement for 5mA beam.}
\end{figure}
\subsection{$^{12}C(p,\gamma)$ REACTION:}
This reaction was carried out with a 50$\mu$g/$cm^2$ graphite foil \citep{burtebaev2008new}. This foil was produced by evaporation of natural carbon onto 1.5-mm-thick Copper backing. The proton current used was 5-15$\mu$A. The energy of the proton beam was 0.5 MeV. As the actual target is very thin compared to the backing material, most of the beam energy is deposited in the Cu backing. So the temperature profile is studied for the backing material. The results are shown in Figure 10(a) and (b) for the case of cooling by radiation only. The maximum temperature of Cu with 15$\mu$A beam current is about $998^oC$ which is below its melting point. From our calculations we see that, with the backing target system used in \citep{burtebaev2008new}, if the water cooling is applied on the backing, the usable current can go upto 3.5mA. This calculations are shown in Figure 11(a) and (b).\\
In (p,$\gamma$) experiments as the outgoing particle is not a charge particle a thick cooled backing behind the target can help using a much higher current.
\begin{figure}
  \centering
  \begin{subfigure}[b]{0.48\linewidth}
    \includegraphics[width=\linewidth]{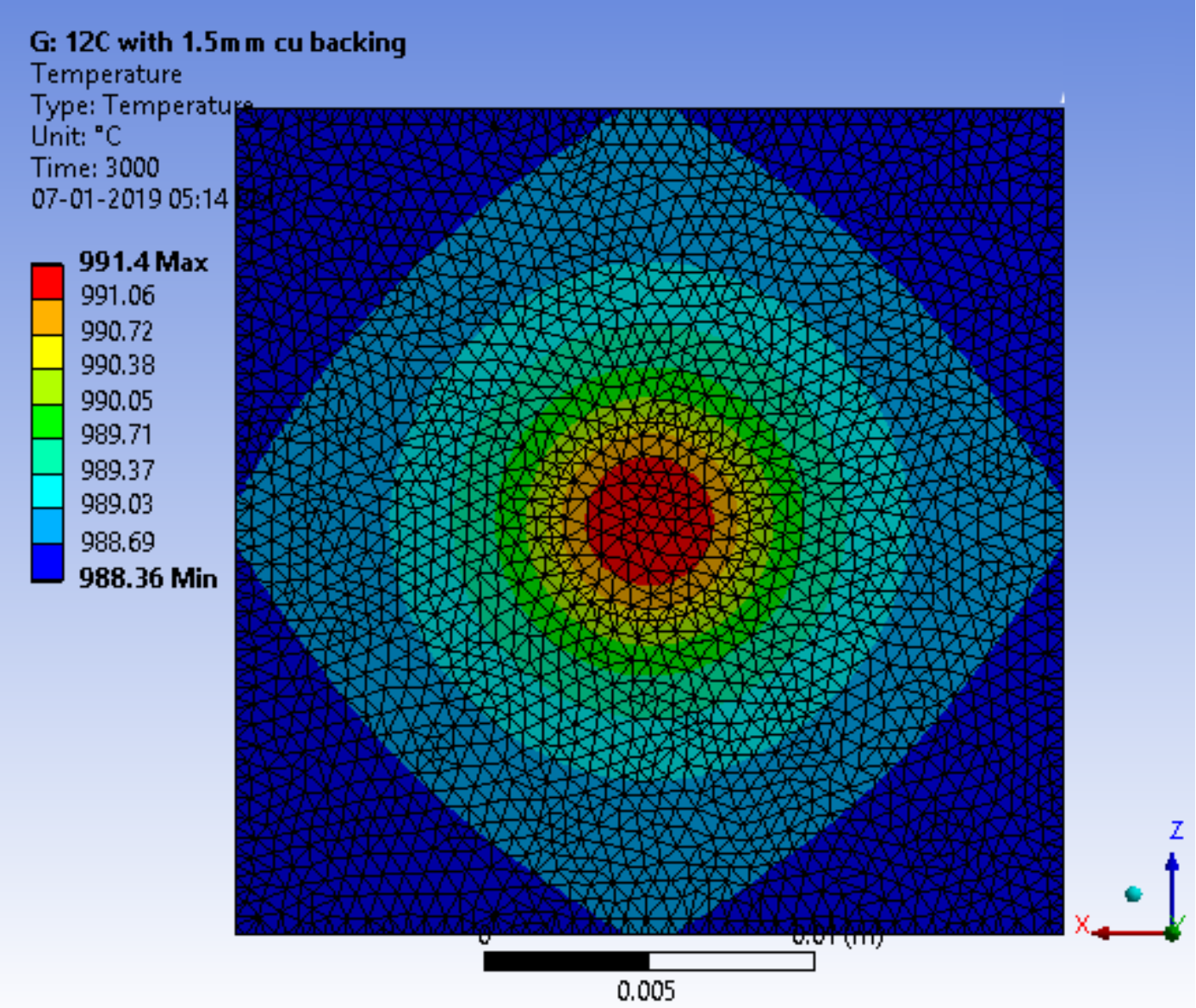}
    \caption{Temperature profile of target.}
  \end{subfigure}
  \begin{subfigure}[b]{0.48\linewidth}
    \includegraphics[width=\linewidth]{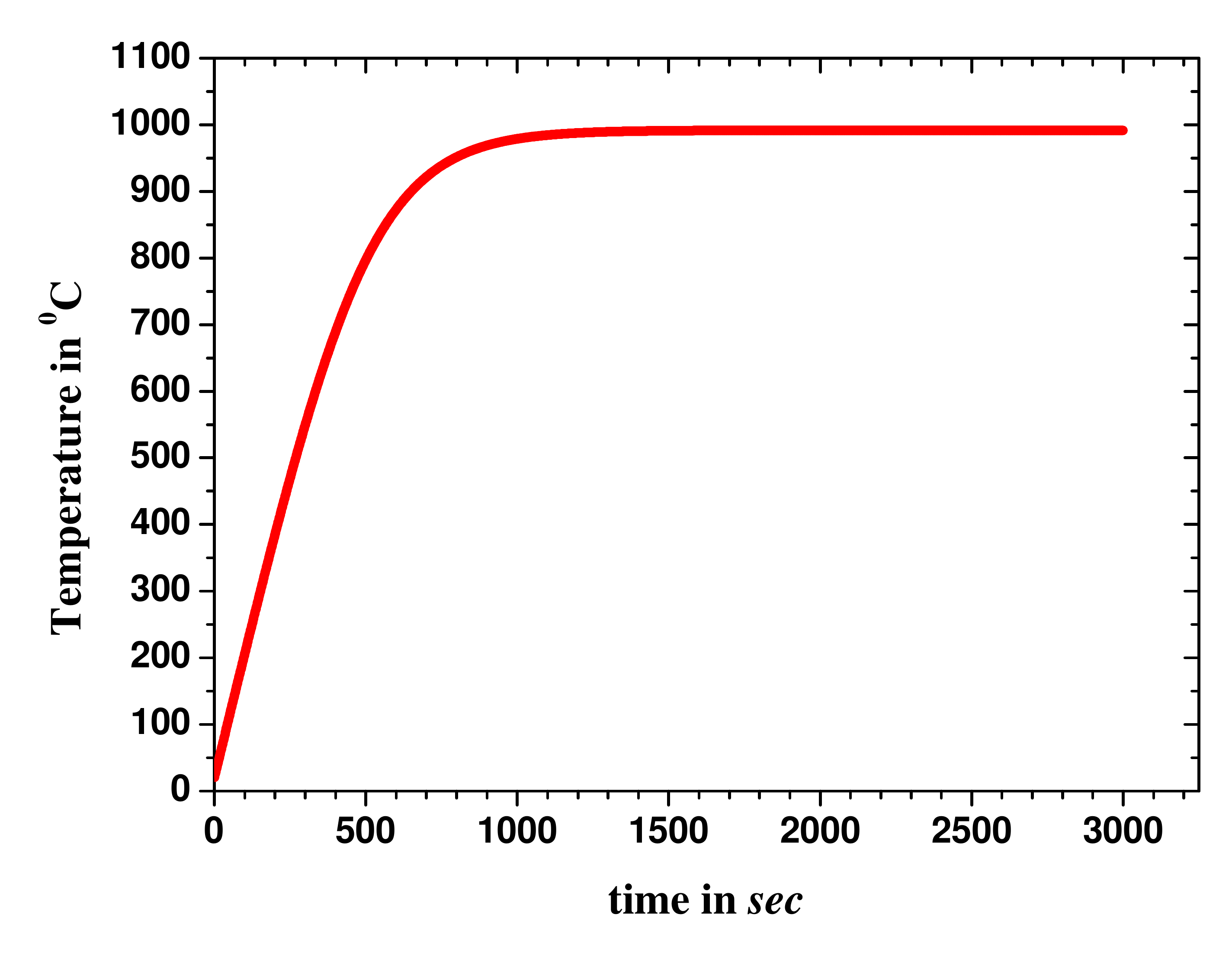}
    \caption{Temporal variation of maximum temperature}
  \end{subfigure}
  \caption{Temperature study of Cu backing without any cooling arrangement for 15$\mu$A beam.}
\end{figure}
\begin{figure}
  \centering
  \begin{subfigure}[b]{0.48\linewidth}
    \includegraphics[width=\linewidth]{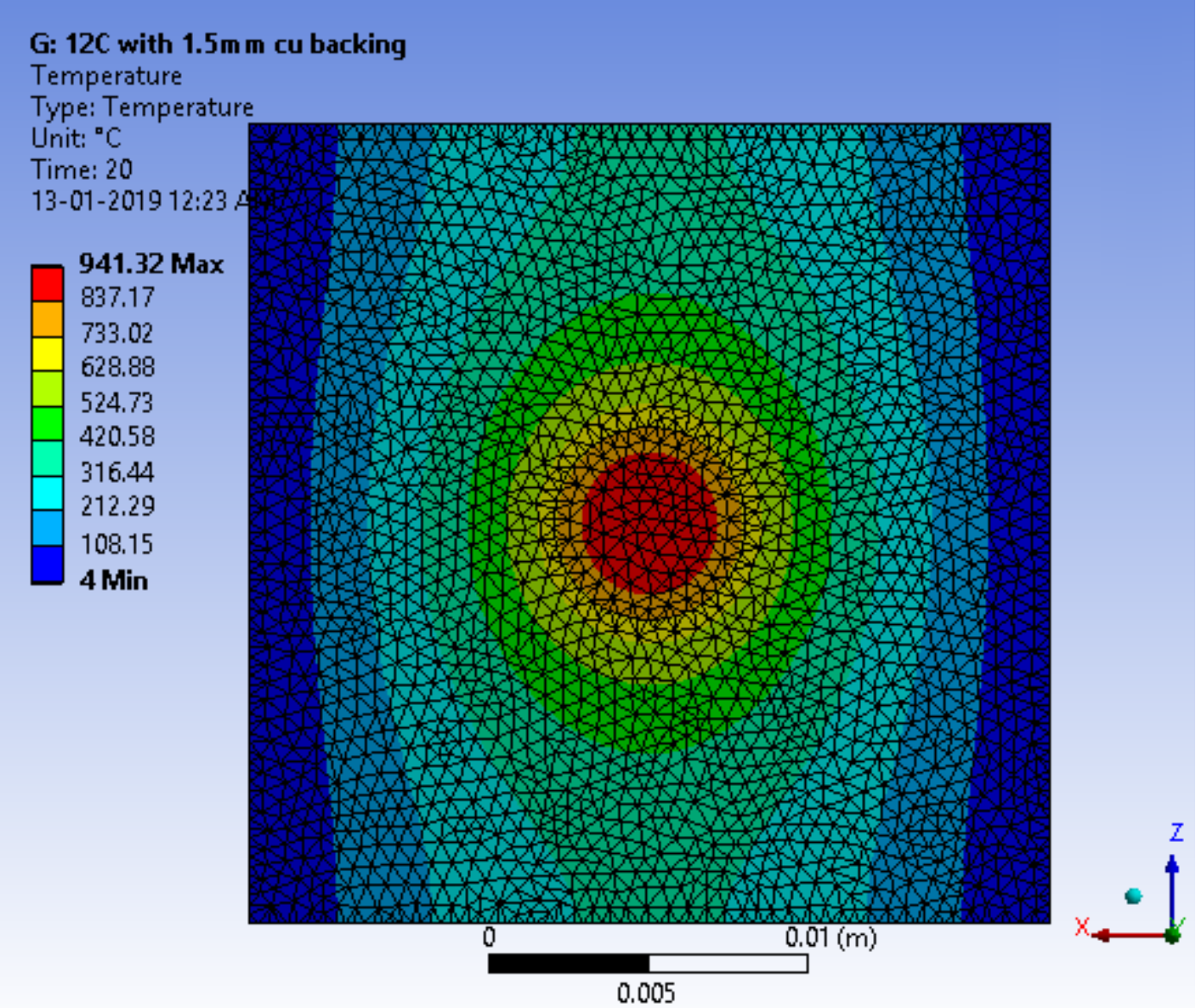}
    \caption{Temperature profile of target.}
  \end{subfigure}
  \begin{subfigure}[b]{0.5\linewidth}
    \includegraphics[width=\linewidth]{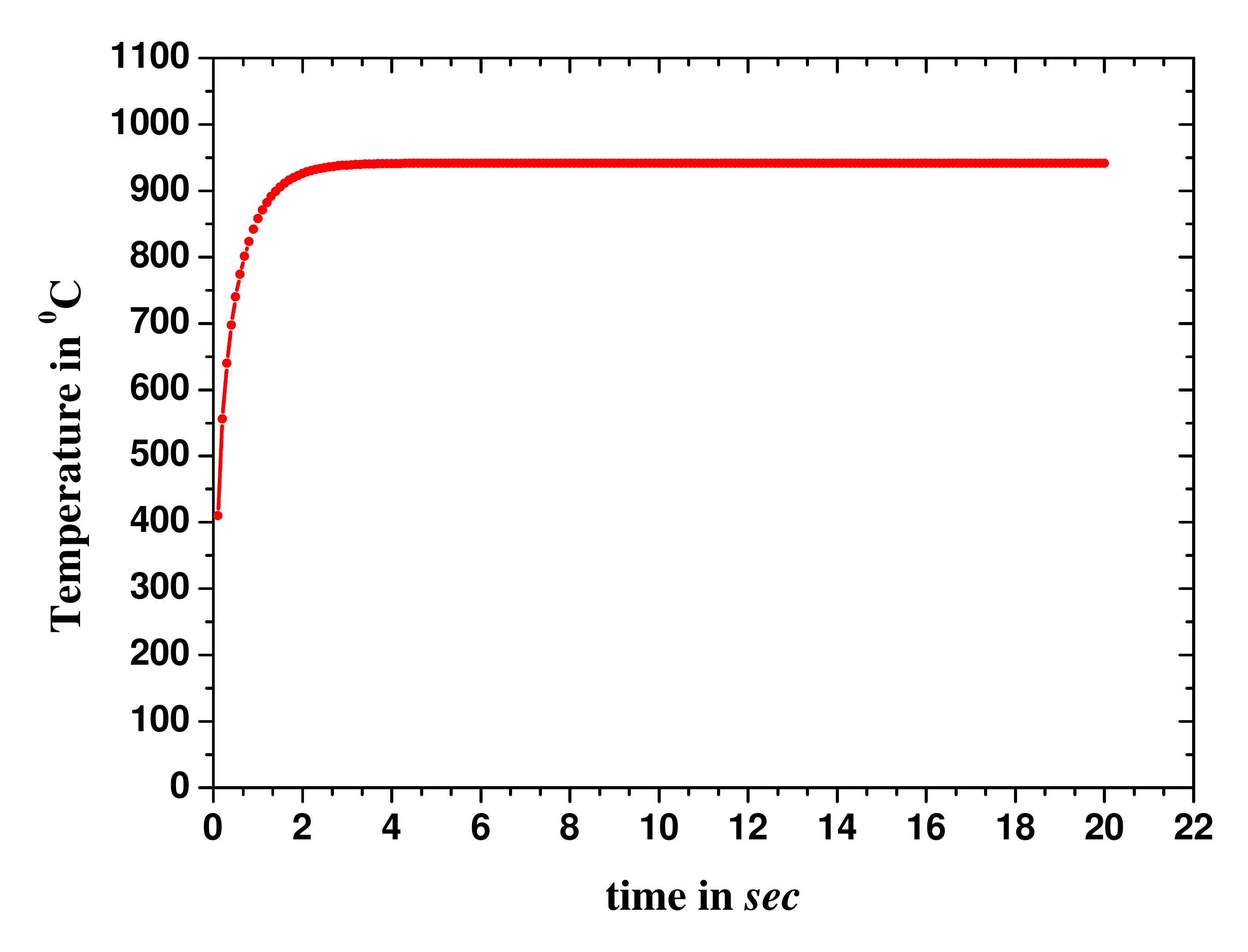}
    \caption{Temporal variation of maximum temperature}
  \end{subfigure}
  \caption{Temperature study of Cu backing with two sides at $4^0$C for 3.5mA beam current.}
\end{figure}
 \subsection{$^{12}C(^{12}C,x)$ REACTION:}
 This reaction was studied where a 1mm ($\sim 225.3mg/cm^2$) graphite target was used \cite{spillane2007c}. The $^{12}C$ beam energy was varied between 4.2-9.5 MeV and the beam ($^{12}C^{2+}$) current was 40$\mu$A [Figure 12(a)-(e)]. The beam current can be raised to 250$\mu$A by considering dissipation by radiation only. These calculations are shown in Figure 13(a)-(e). The maximum temperature is around $3700^oC$.\\
 If now the chilled water cooling is applied on the two sides of the target the maximum temperature and beam current is not much altered [Figure 14(a)-(e)].This happened due to the very low conductivity of carbon. For back side $4^0$C cooling arrangement the maximum beam current increases to 1mA for $^{12}C^{2+}$ beam [Fig.15(a) and (c)].

 \begin{figure}
  \centering
  \begin{subfigure}[b]{0.45\linewidth}
    \includegraphics[width=\linewidth]{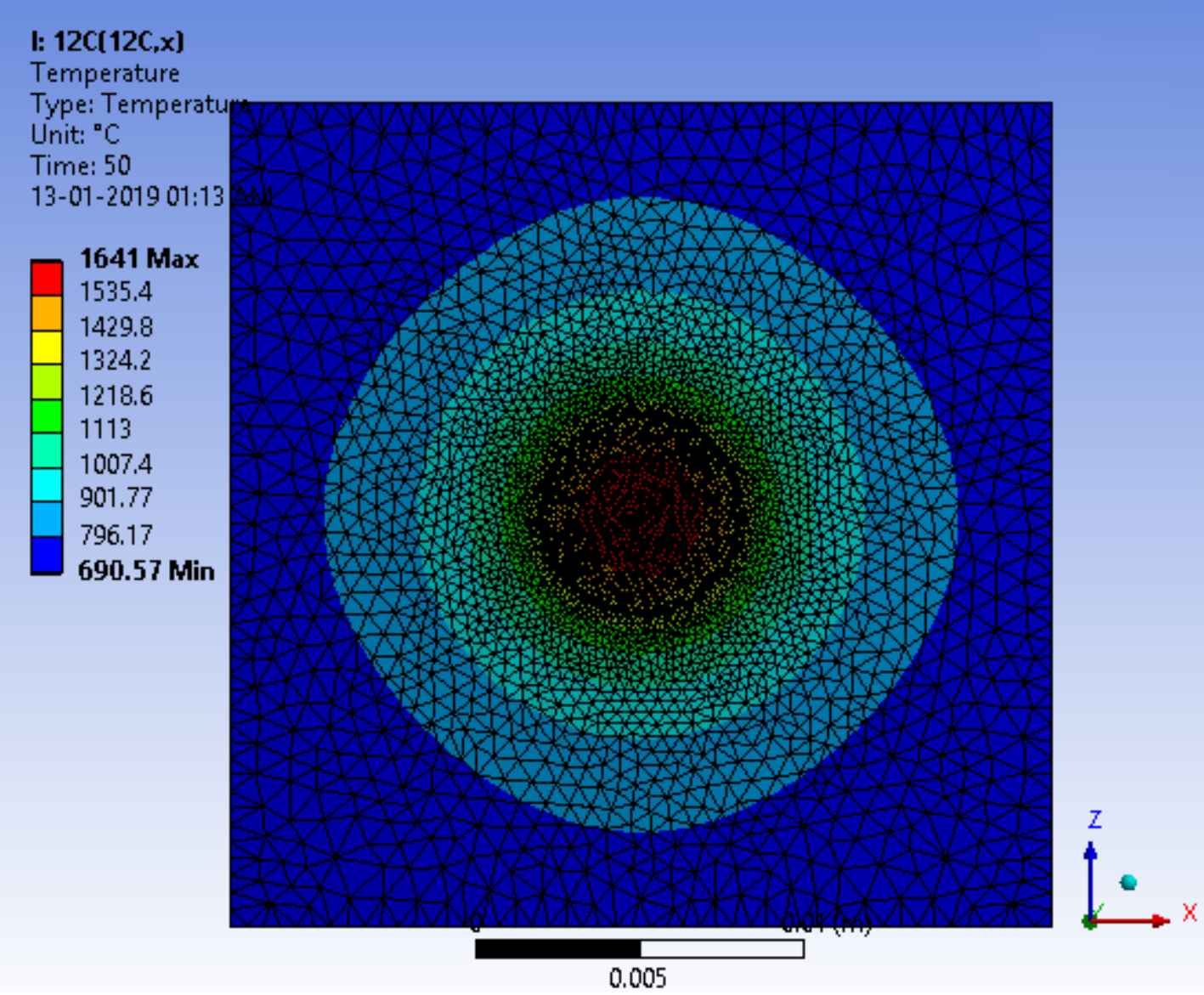}
    \caption{Front side temperature profile of target.}
  \end{subfigure}
  \begin{subfigure}[b]{0.45\linewidth}
    \includegraphics[width=\linewidth]{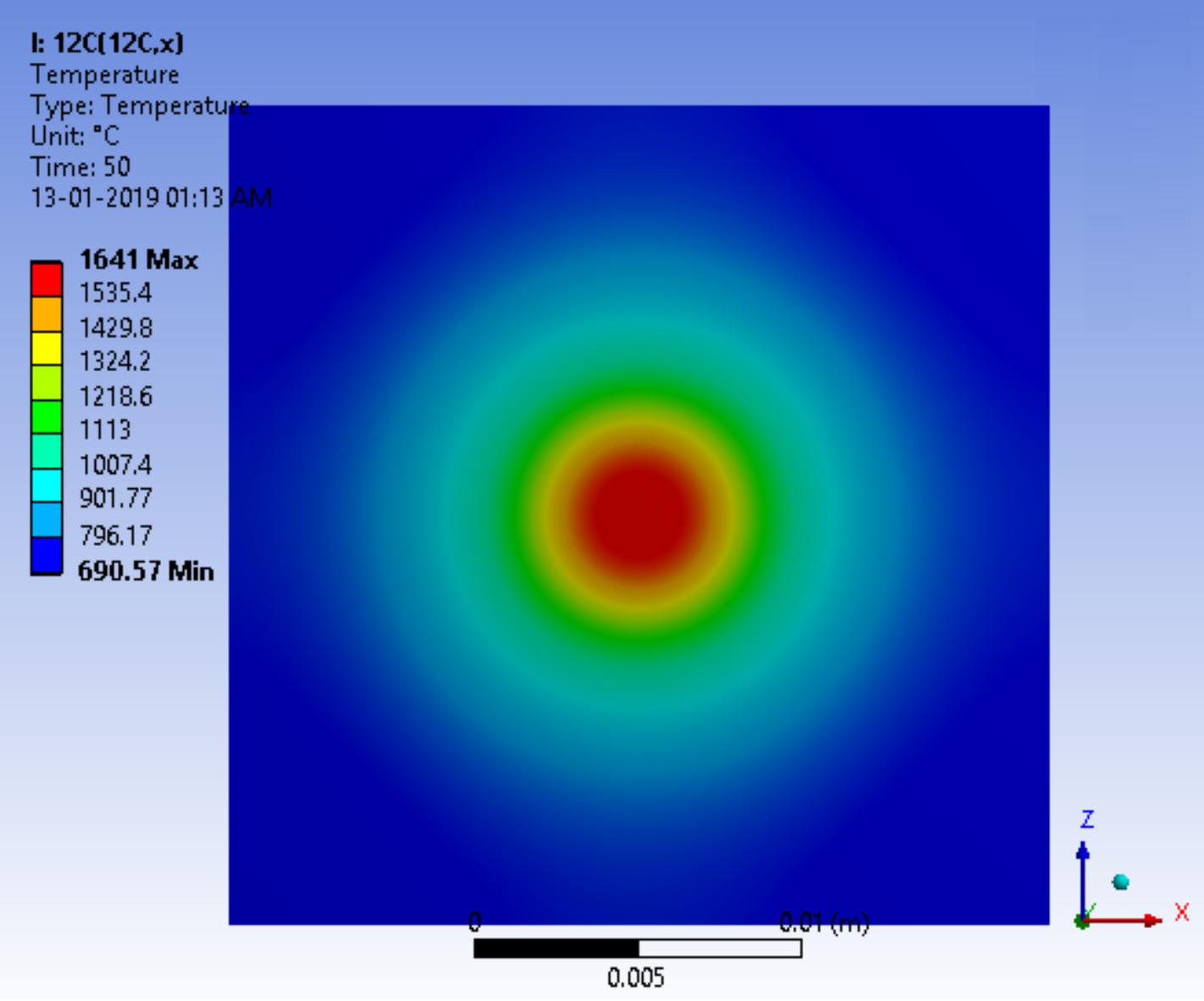}
    \caption{Front side temperature profile without grids.}
  \end{subfigure}
  \begin{subfigure}[b]{0.45\linewidth}
    \includegraphics[width=\linewidth]{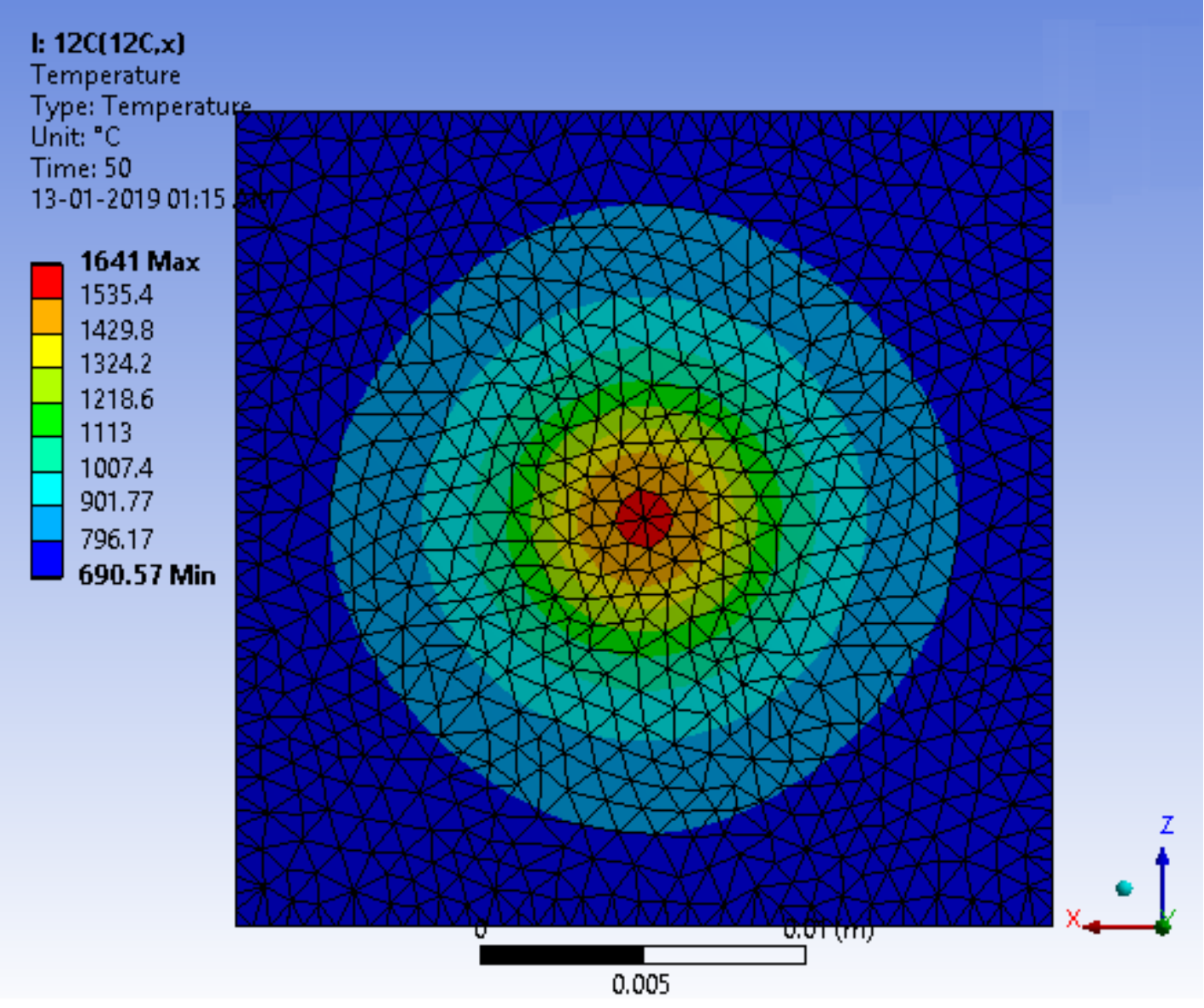}
    \caption{Back side temperature profile of target}
    \end{subfigure}
  \begin{subfigure}[b]{0.45\linewidth}
    \includegraphics[width=\linewidth]{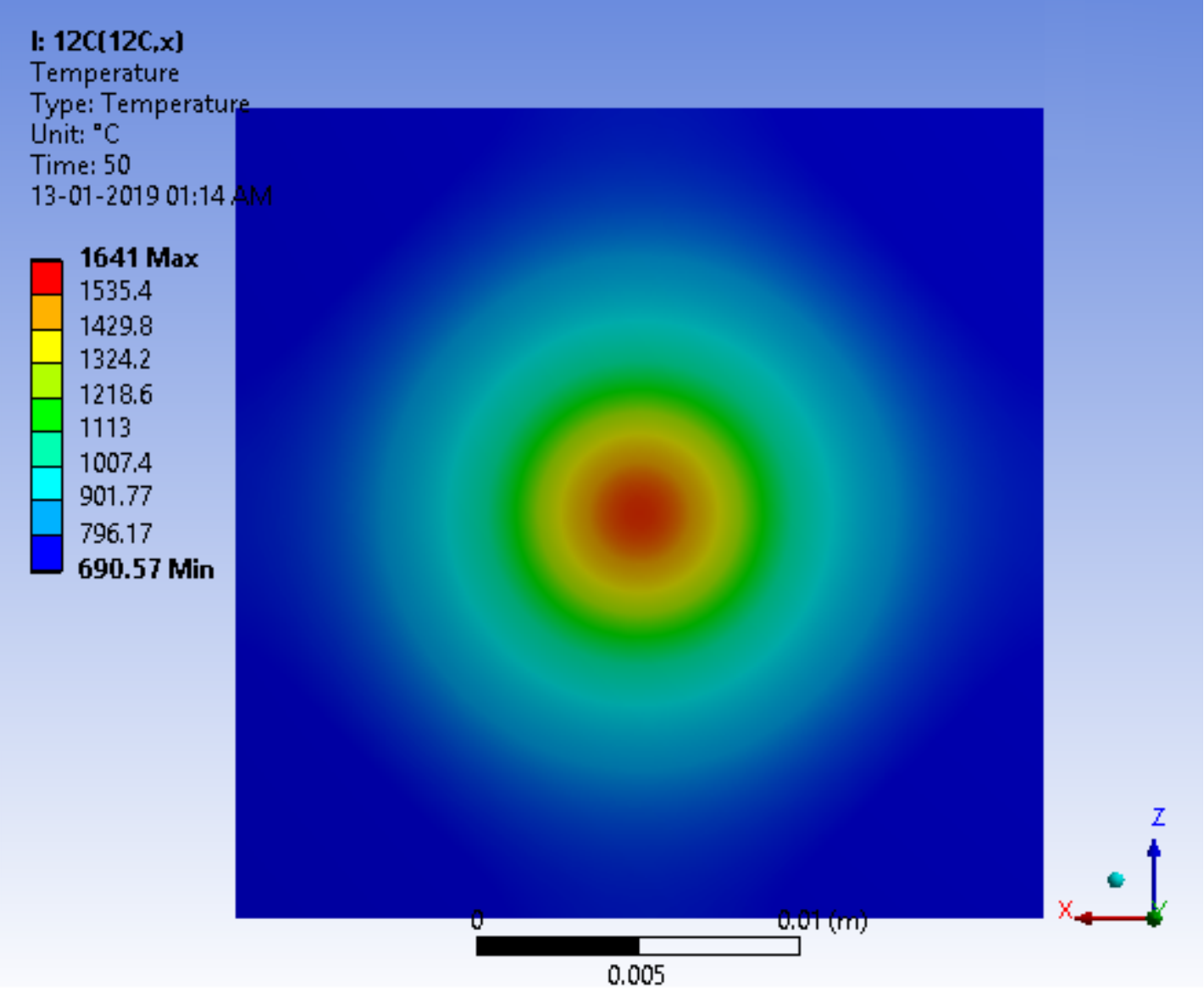}
    \caption{Back side temperature profile without grids.}
    \end{subfigure}
    \begin{subfigure}[b]{0.5\linewidth}
    \includegraphics[width=\linewidth]{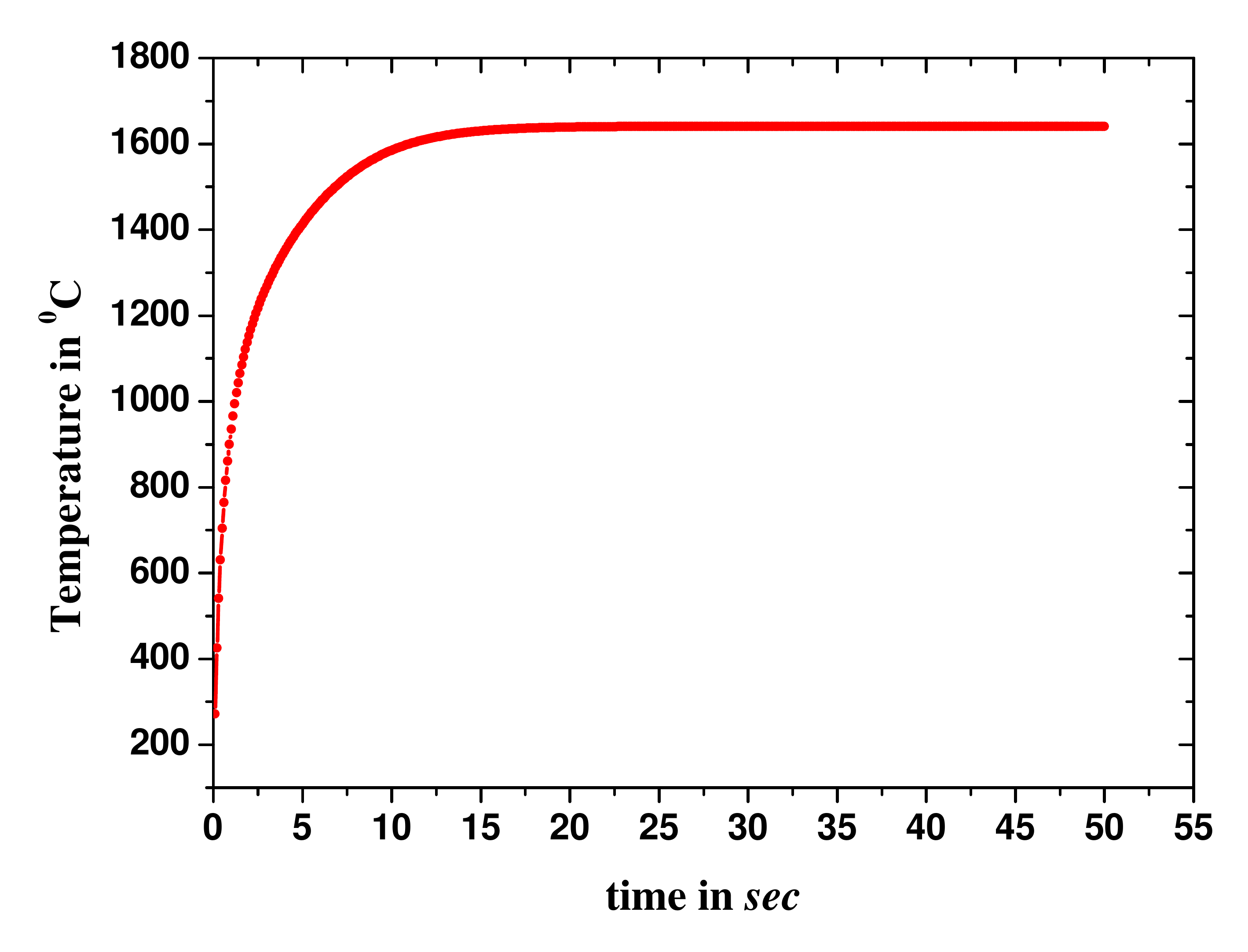}
    \caption{Temporal variation of maximum temperature.}
  \end{subfigure}
  \caption{Temperature study of $^{12}C(^{12}C,x)$ reaction without any cooling arrangement for 40 \textit{$\mu$A} beam.}
\end{figure}

 \begin{figure}
  \centering
  \begin{subfigure}[b]{0.45\linewidth}
    \includegraphics[width=\linewidth]{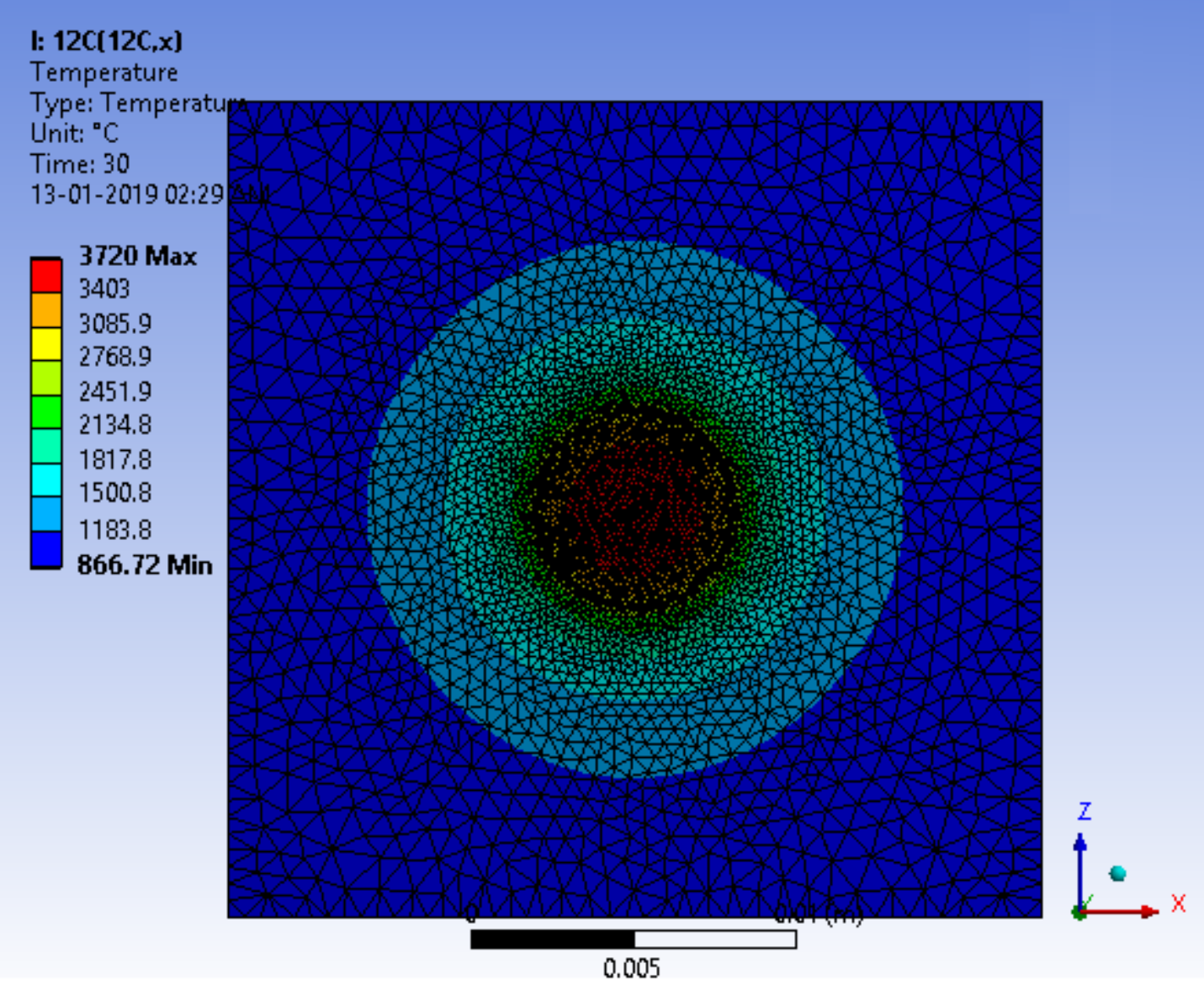}
    \caption{Front side temperature profile of target.}
  \end{subfigure}
  \begin{subfigure}[b]{0.44\linewidth}
    \includegraphics[width=\linewidth]{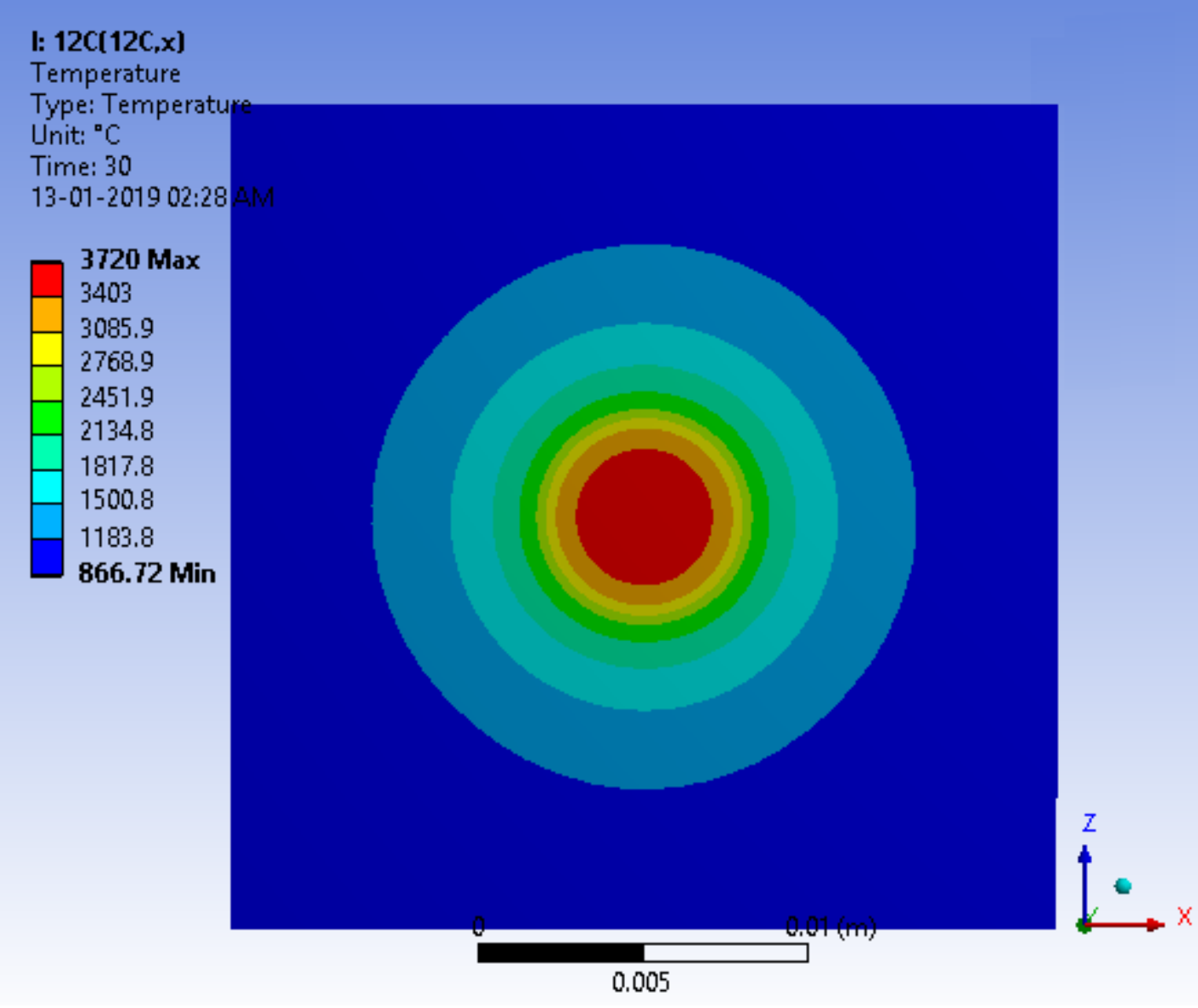}
    \caption{Front view of target without grids.}
  \end{subfigure}
  \begin{subfigure}[b]{0.45\linewidth}
    \includegraphics[width=\linewidth]{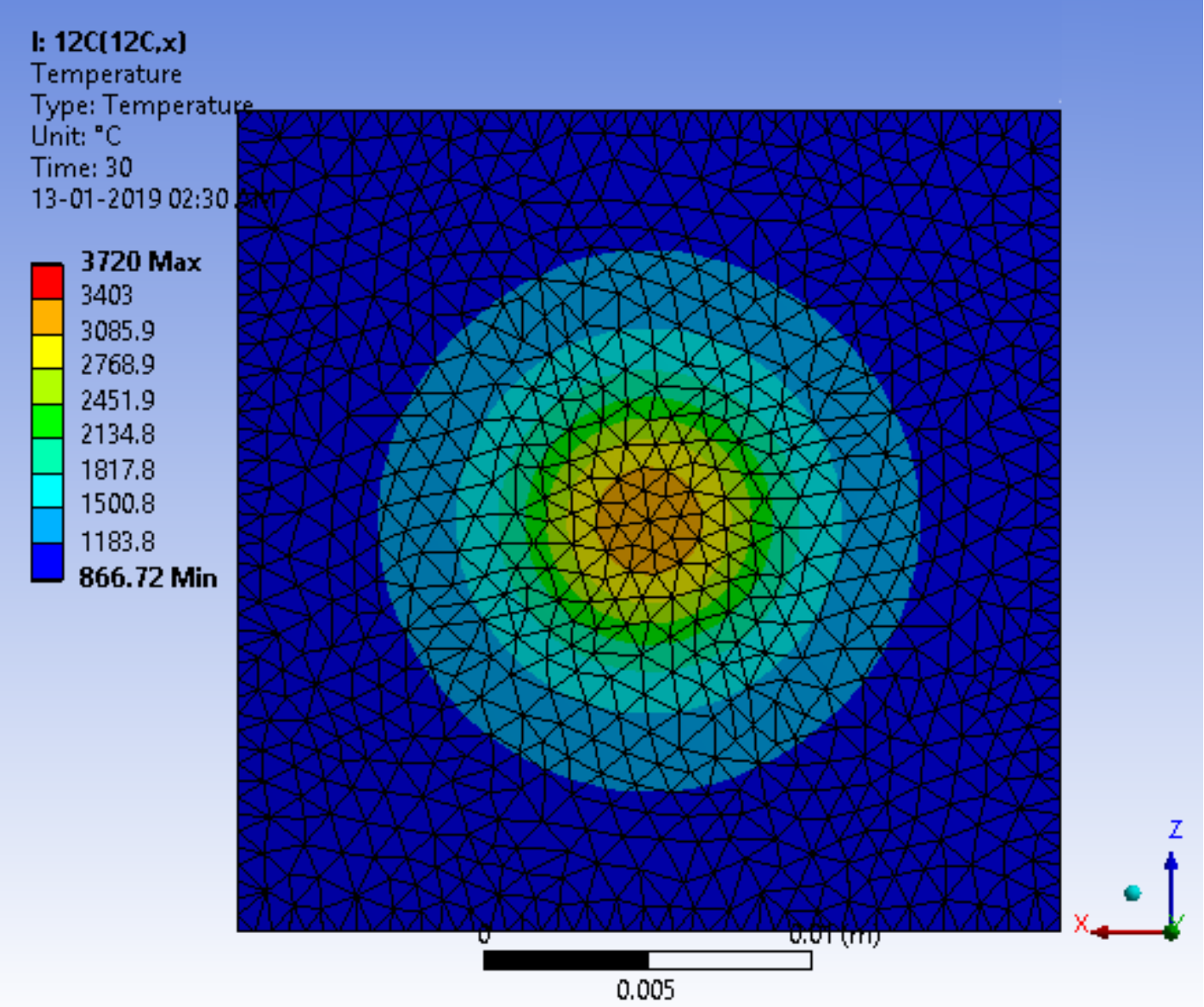}
    \caption{Back side temperature profile of target}
    \end{subfigure}
  \begin{subfigure}[b]{0.45\linewidth}
    \includegraphics[width=\linewidth]{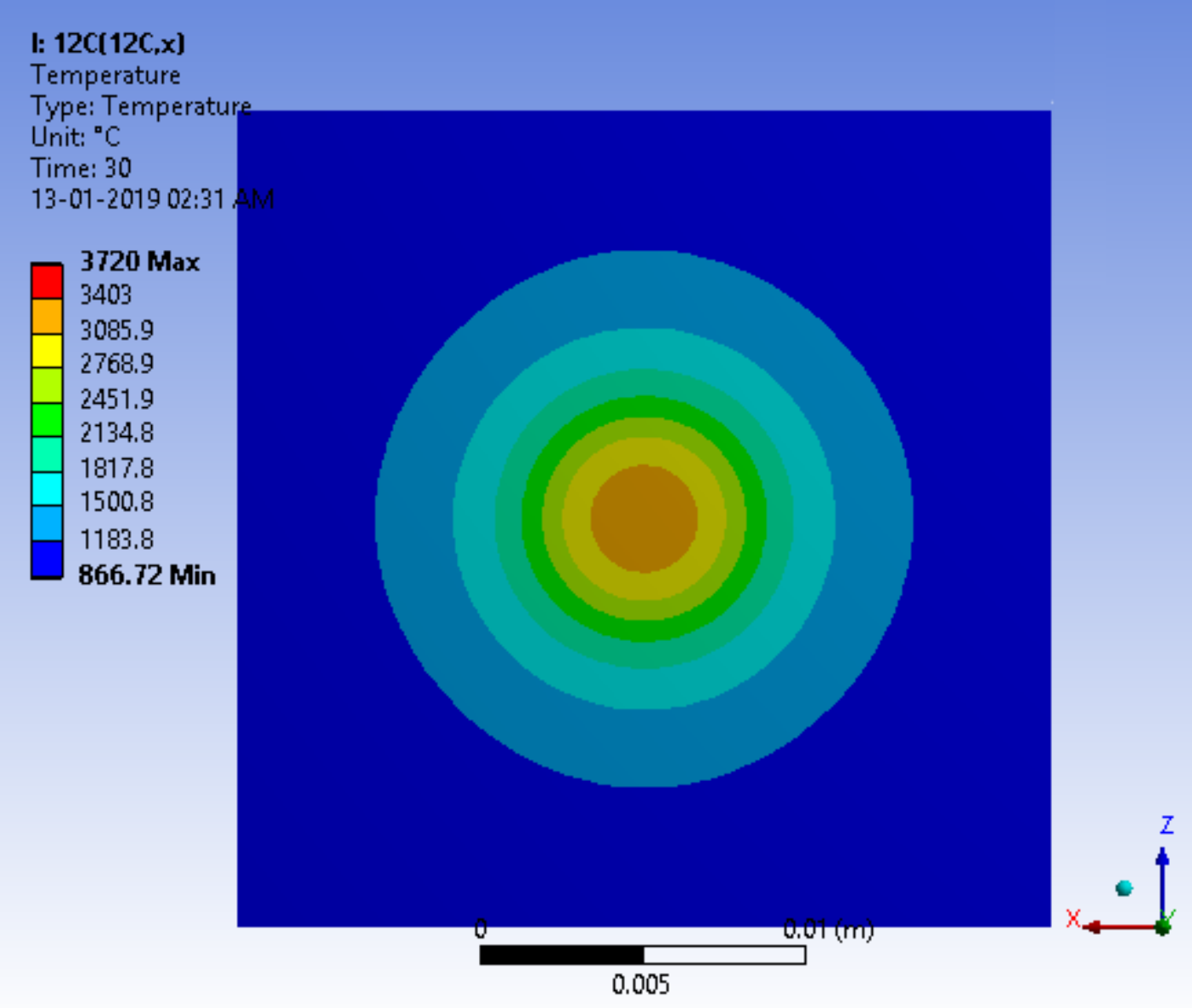}
    \caption{Back side temperature profile without grids}
  \end{subfigure}
  \begin{subfigure}[b]{0.5\linewidth}
    \includegraphics[width=\linewidth]{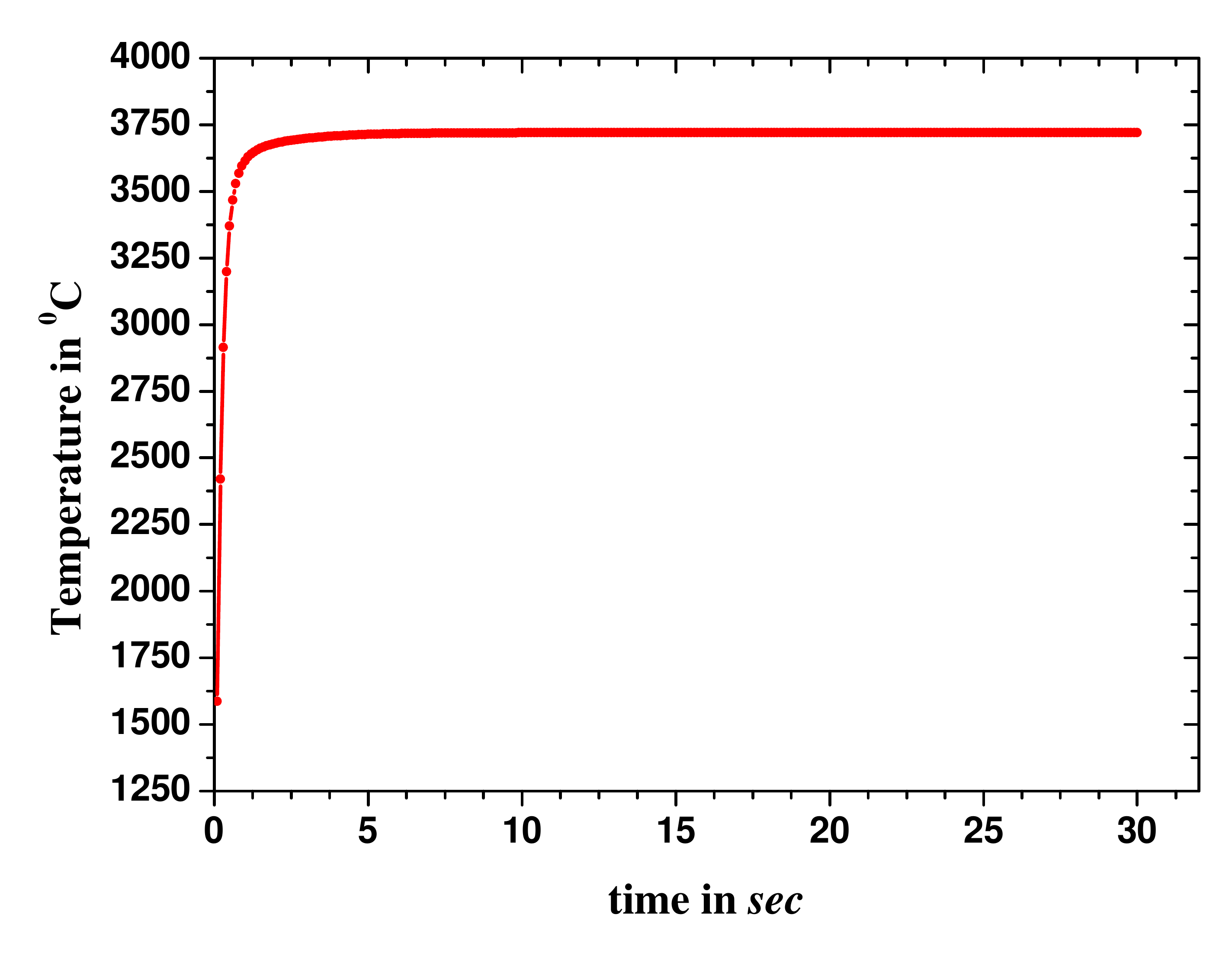}
    \caption{Temporal variation of maximum temperature}
  \end{subfigure}
  \caption{Temperature study of $^{12}C(^{12}C,x)$ reaction without any cooling arrangement for 250 \textit{$\mu$A} beam.}
\end{figure}
 
 \begin{figure}
  \centering
  \begin{subfigure}[b]{0.45\linewidth}
    \includegraphics[width=\linewidth]{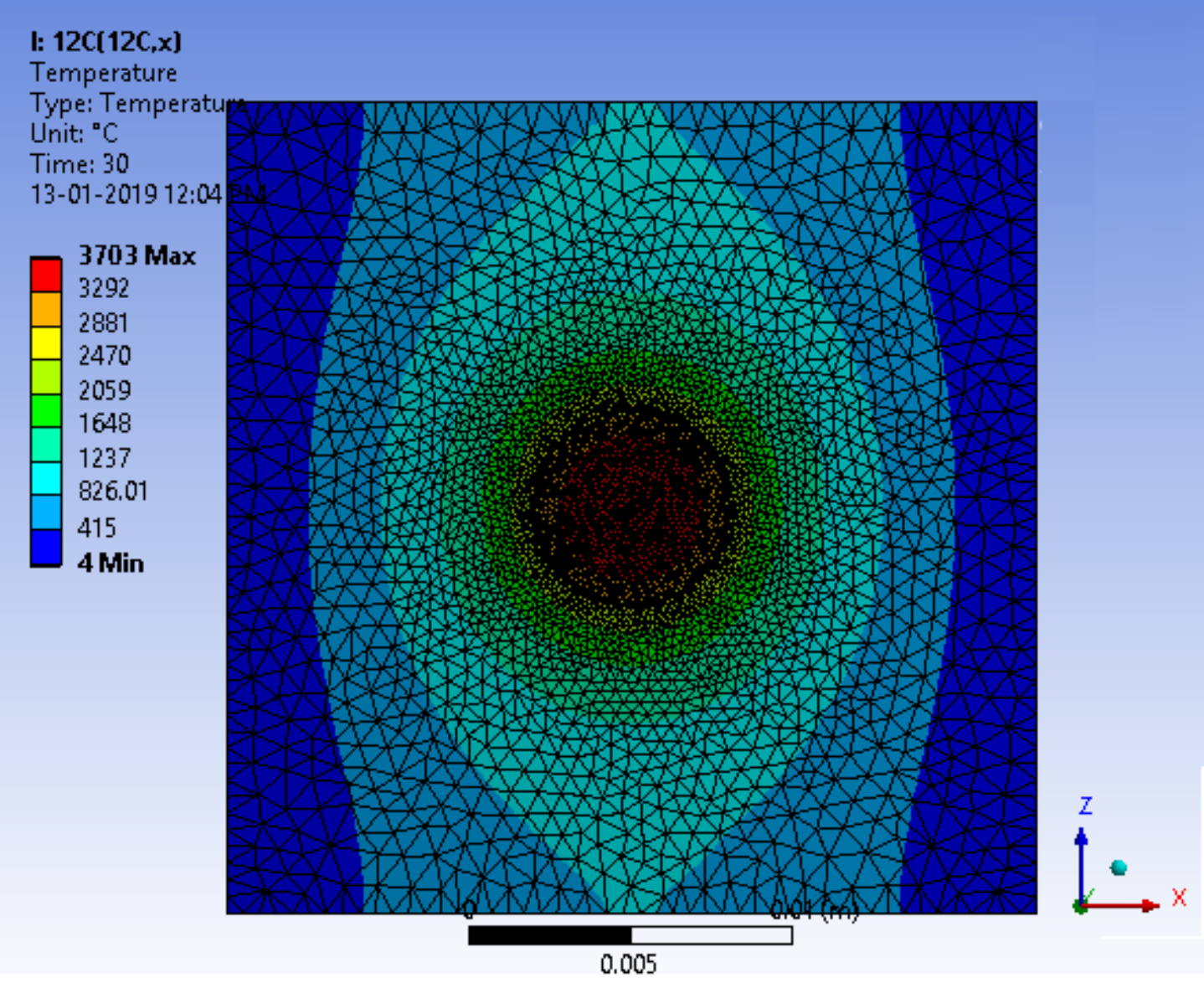}
    \caption{Front side temperature profile of target}
  \end{subfigure}
  \begin{subfigure}[b]{0.45\linewidth}
    \includegraphics[width=\linewidth]{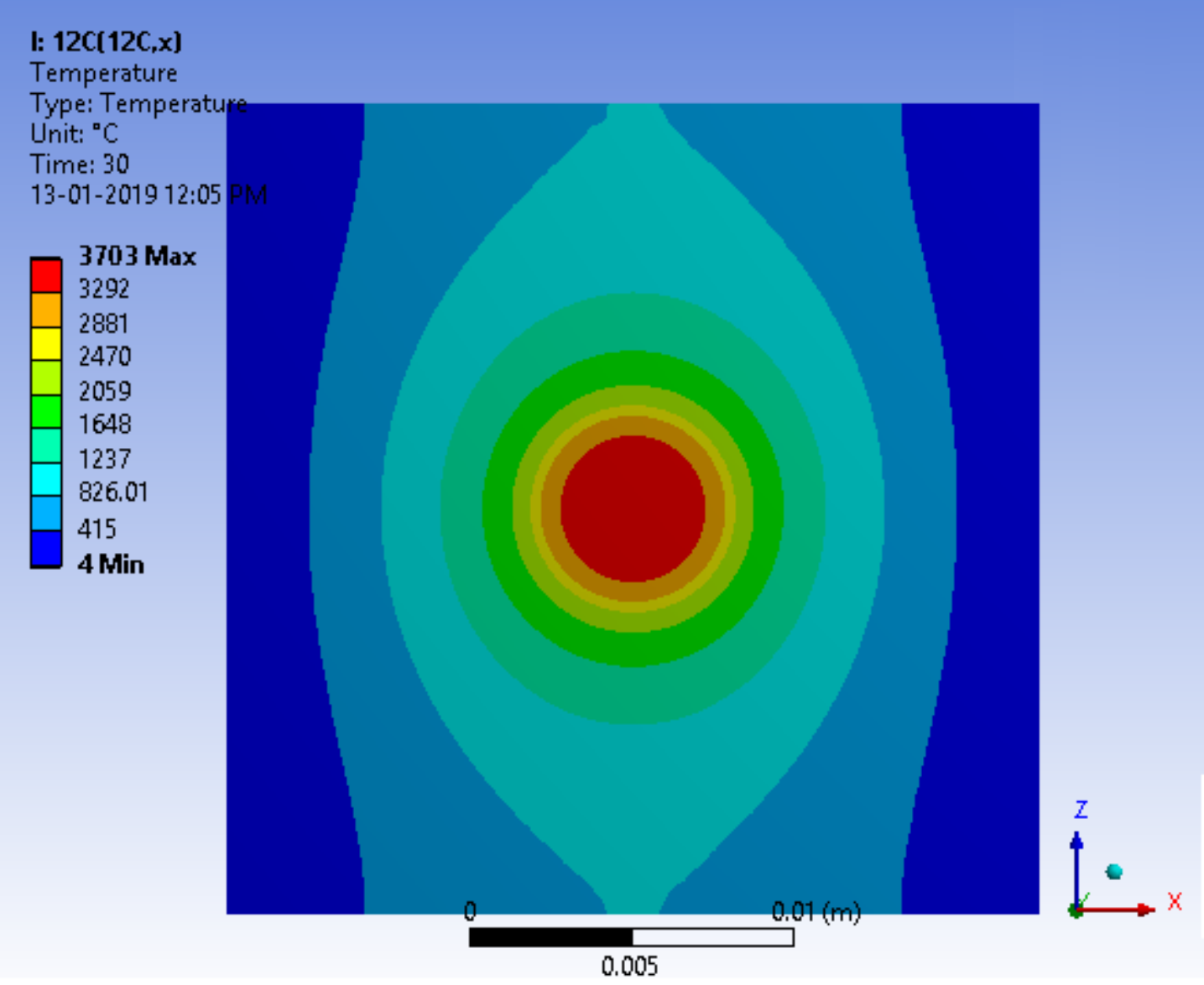}
    \caption{Front side temperature profile without grids.}
    \end{subfigure}
  \begin{subfigure}[b]{0.45\linewidth}
    \includegraphics[width=\linewidth]{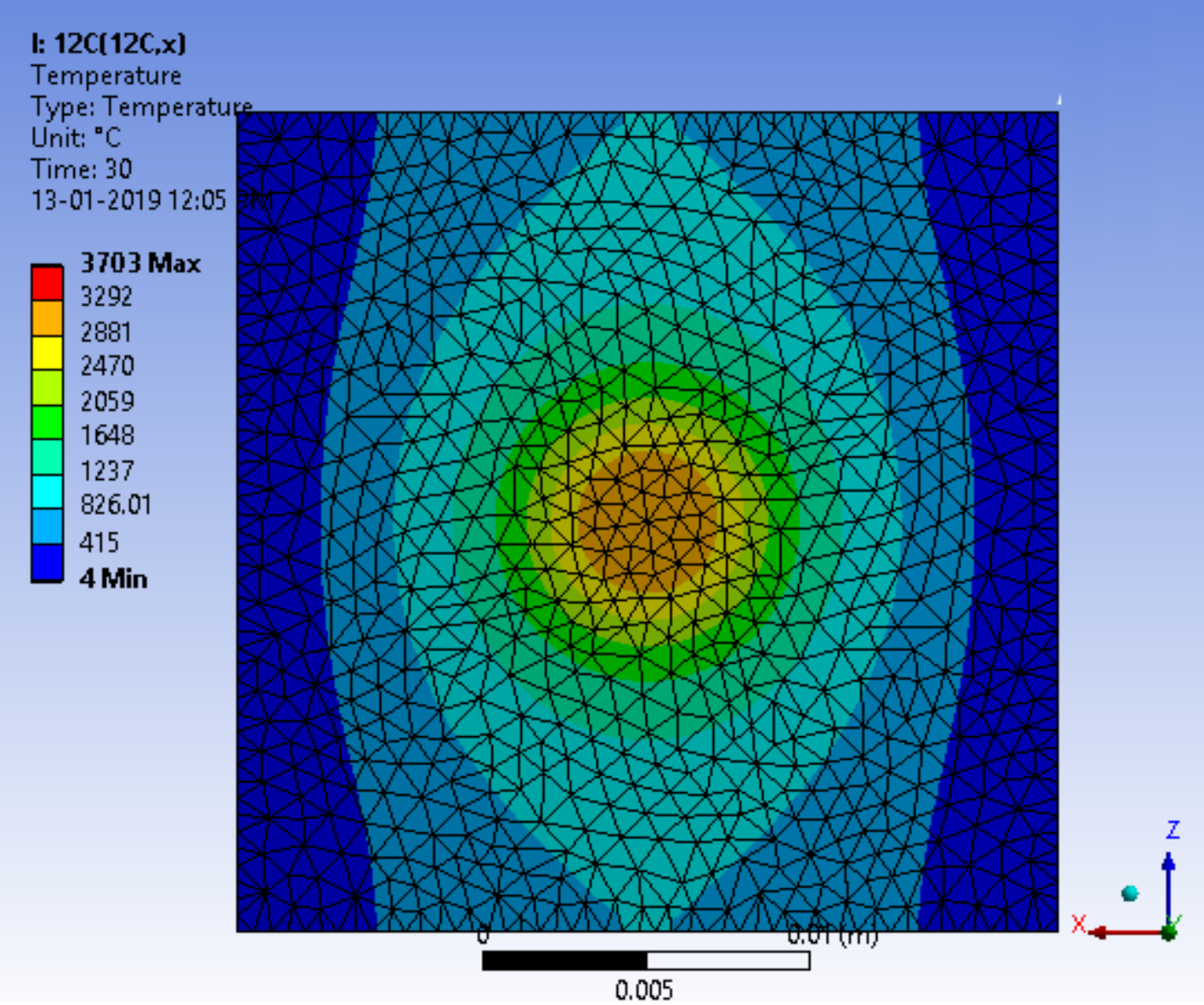}
    \caption{Back side temperature profile of target}
  \end{subfigure}    
  \begin{subfigure}[b]{0.45\linewidth}
    \includegraphics[width=\linewidth]{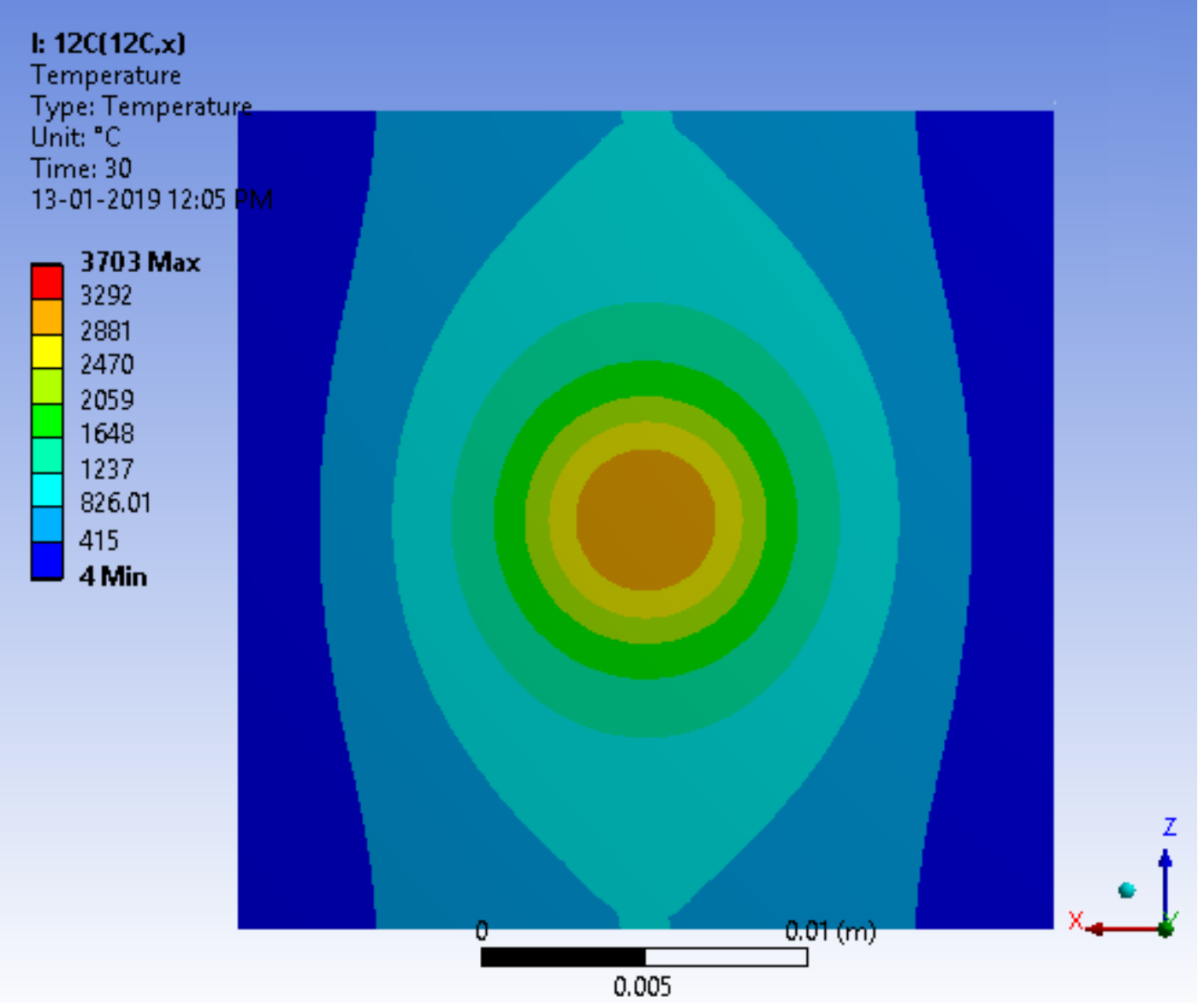}
    \caption{Back side temperature profile without grids.}
  \end{subfigure}
    \begin{subfigure}[b]{0.5\linewidth}
    \includegraphics[width=\linewidth]{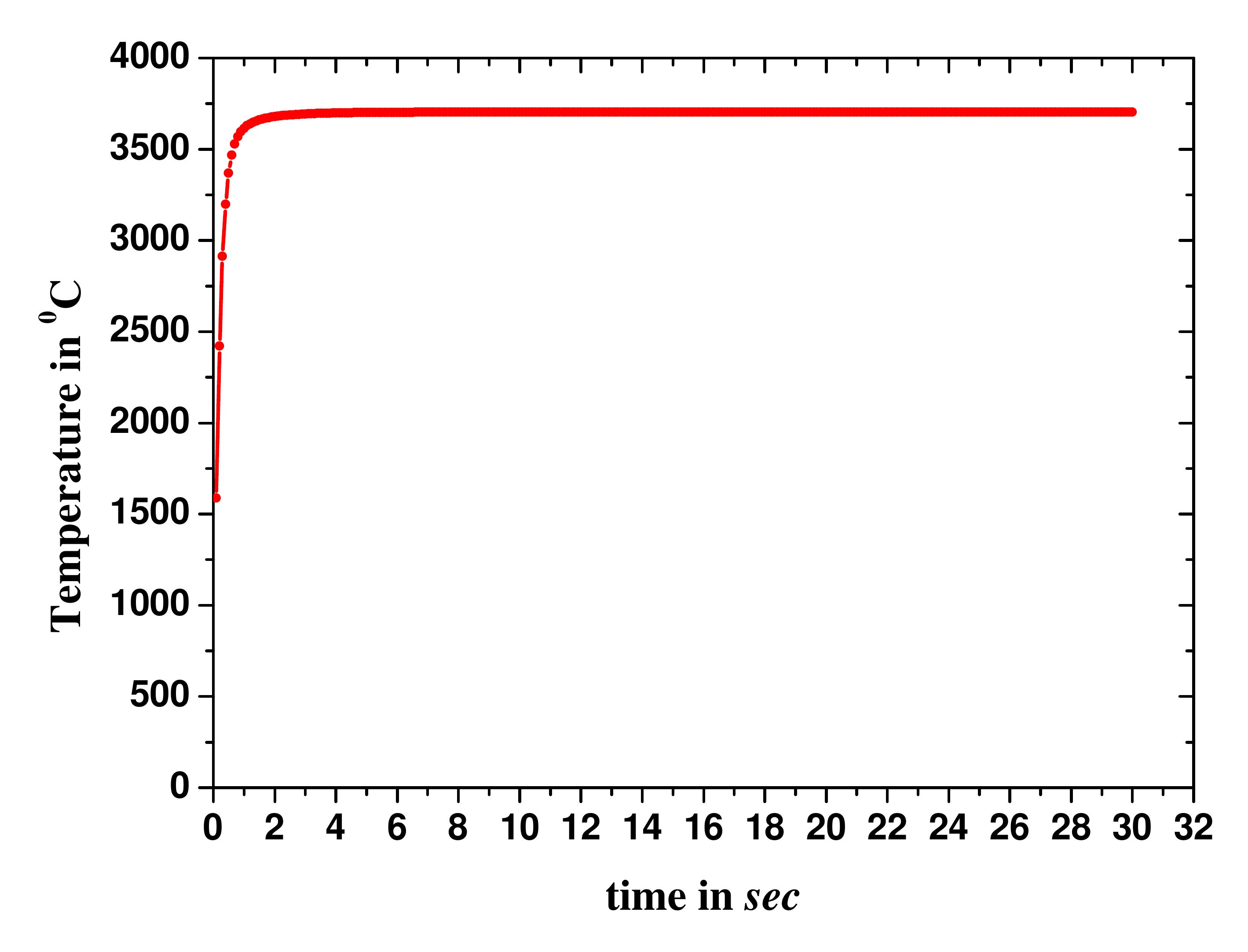}
    \caption{Temporal variation of maximum temperature.}
  \end{subfigure}
  \caption{Temperature study of $^{12}C(^{12}C,x)$ reaction with two side cooling arrangement for 250 \textit{$\mu$A} beam.}
\end{figure}

 \begin{figure}
  \centering
  \begin{subfigure}[b]{0.45\linewidth}
    \includegraphics[width=2.4 in]{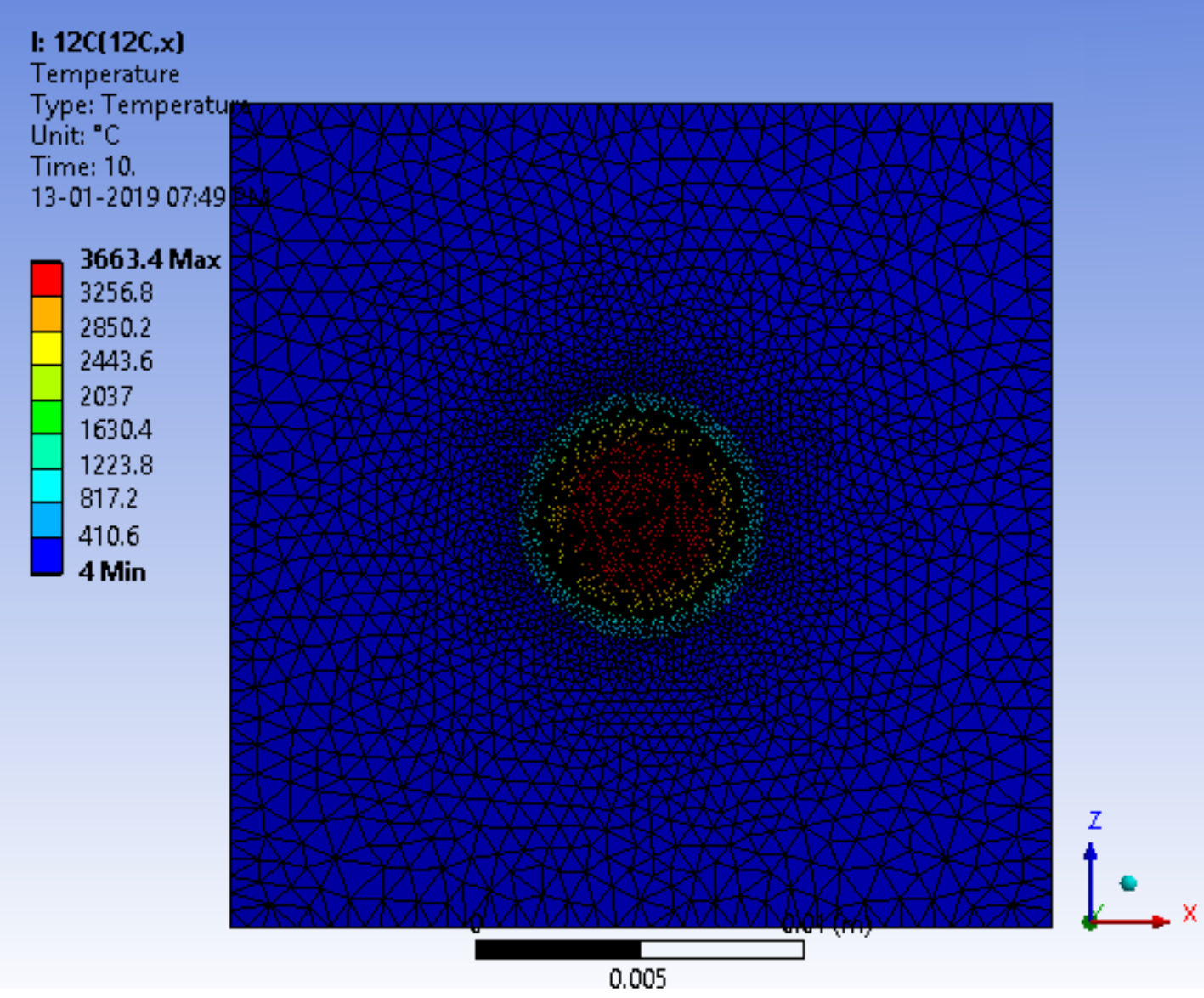}
    \caption{Front side temperature profile of target.}
  \end{subfigure}
    \begin{subfigure}[b]{0.45\linewidth}
    \includegraphics[width=2.4 in]{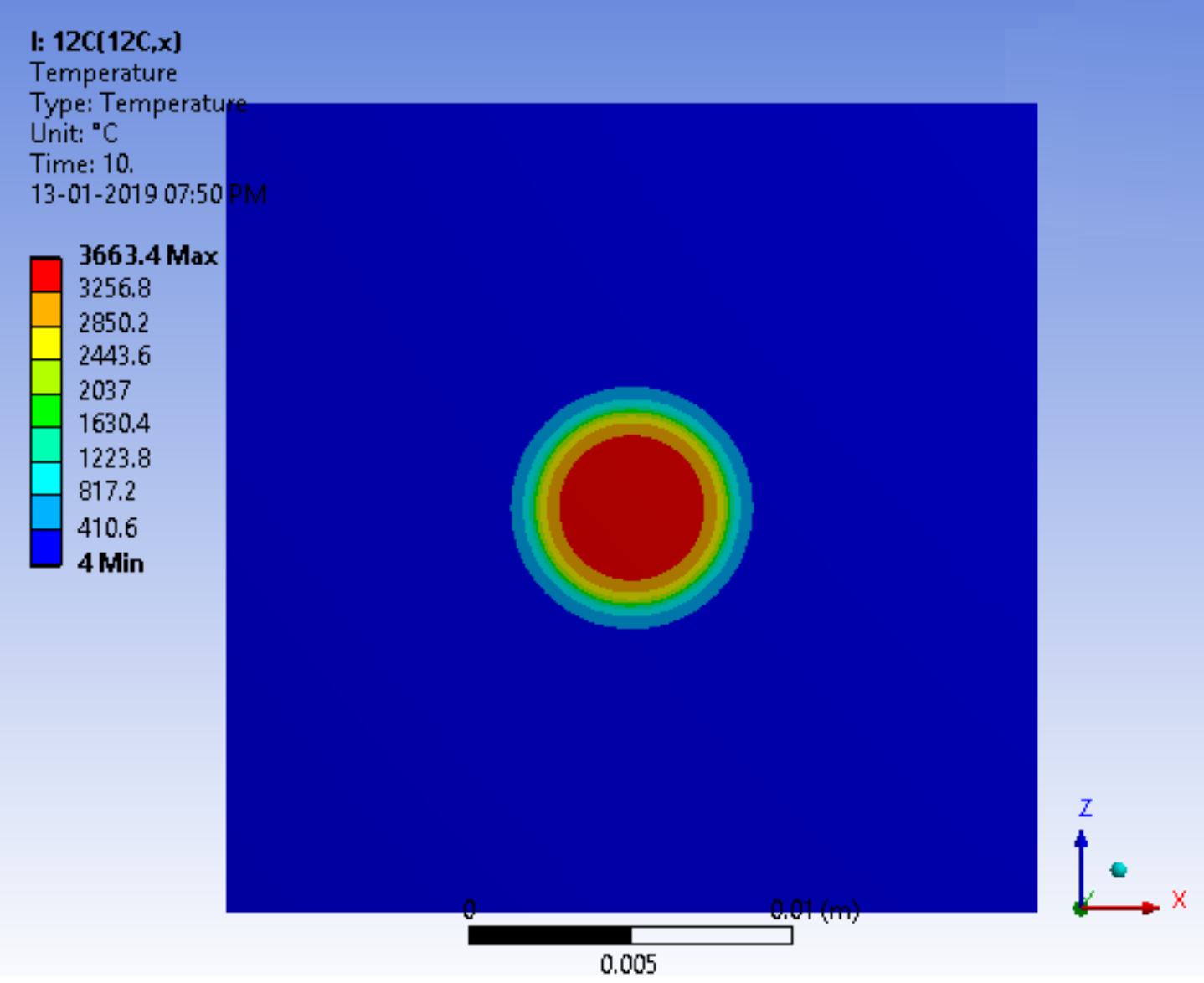}
    \caption{Front side temperature profile without grids.}
  \end{subfigure}
  \begin{subfigure}[b]{0.5\linewidth}
    \includegraphics[width=\linewidth]{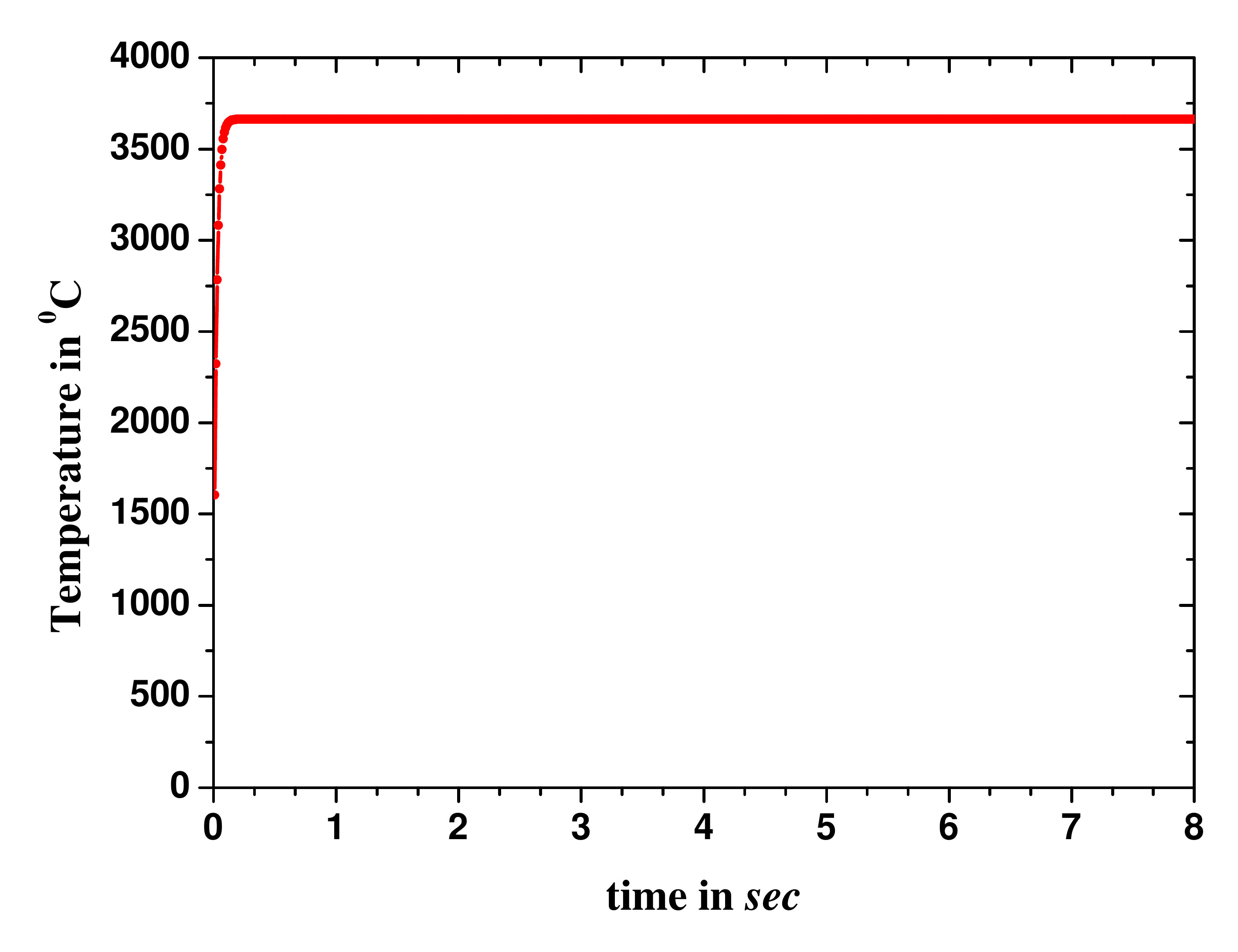}
    \caption{Temporal variation of maximum temperature}
  \end{subfigure}
  \caption{Temperature study of $^{12}C(^{12}C,x)$ reaction with back side cooling arrangement for 1mA beam.}
\end{figure}

\clearpage
\section{Summary and conclusion}
In this work, we have performed a numerical study on the heating of nuclear targets in high current ion-beam experiments. Targets of low and high thermal conductivity viz. $^{27}Al$ and $^{12}C$ have been considered for the calculations. The thickness, ion-beam current was taken from the published experiments as starting points. ANSYS software which solves a time dependent heat dissipation equation is used for the simulation.\\
The results show that for self supporting thin targets(normally used for charge particle emitting reactions) the heat generation is quite low and quite large currents upto a few tens of microamperes can be conveniently used for targets of both low and high thermal conductvities. In case of capture reaction thin targets on an appropriate thick backing which has good thermal conductivity may be used. This backing target can be cooled to achieve higher currents. In case of heavy ion collisions like $^{12}C+^{12}C$ reactions cooling the back side of thick target is very beneficial to dissipate the heat. However use of thick target will be limited to $\gamma$-methods only. In case of thin carbon targets no cooling is required.
\section*{Acknowledgement}
We would like to thank Power Engineering Department of Jadavpur University, Saltlake campus for providing the opportunity to use their lab and computational facilities.

\end{document}